\documentclass[a4paper,11pt]{article}
\pdfoutput=1 

\usepackage{jheppub} 

\usepackage[T1]{fontenc} 
\usepackage{blindtext}
\usepackage{epstopdf}
\usepackage{graphicx}
\usepackage{epsfig}
\usepackage{dcolumn}  
\usepackage{bm}    
\usepackage{caption}
\usepackage{subcaption}
\usepackage{amssymb} 
\usepackage{tikz}
\usetikzlibrary{arrows}
\usepackage{nccmath}
\usepackage{xcolor}
\usepackage{epstopdf}
\usepackage{graphicx,wrapfig,lipsum}
\usepackage{epsfig}
\usepackage{amsmath,bm}
\usepackage{amsfonts}  
\usepackage{amsmath}  
\usepackage{slashed}  
\usepackage{enumitem}
\usepackage[mathscr]{euscript}
\usepackage{tabu}
\usepackage{epsfig}
\makeatletter
\g@addto@macro\bfseries{\boldmath}
\makeatother

\DeclareMathOperator{\csch}{csch}

\def\l1{{{1-loop}}}

\def\n1{\Bigg|_{n=1}}

\def\n{{(n)}}

\usepackage[T1]{fontenc} 
\usepackage{tikz}
\usepackage{amsmath,amssymb}
\usepackage{relsize}
\usepackage{latexsym}
\usepackage{leftidx}
\usepackage{xcolor}
\usepackage{diagbox}
\usepackage[T1]{fontenc}
\usepackage{array}
\usepackage{makecell}
\usepackage{csquotes}
\usepackage{tikz}
\usepackage{enumitem}
\usepackage{setspace}
\usepackage{multirow}
\usepackage{amsmath,amssymb}
\usepackage{relsize}
\usepackage{latexsym}
\usepackage{leftidx}
\usepackage{xcolor}
\usepackage{csquotes}
\usepackage{tikz}
\usetikzlibrary{decorations.pathmorphing}
\usetikzlibrary{arrows.meta}
\usepackage{enumitem}
\usetikzlibrary{decorations.markings}
\usetikzlibrary{decorations.pathmorphing}
\usetikzlibrary{decorations.markings}
\usetikzlibrary{decorations.pathmorphing}
\usepackage{pifont}
\usepackage{hyperref}
\usepackage{url}

\usepackage{bookmark}

\title{\textbf{\textsf{The  large $N$ vector model 
on $S^1\times S^2$ 
}}}
\author{Justin R. David, Srijan Kumar}
\affiliation{\vspace{.1cm} Centre for High Energy Physics, \\ Indian Institute of Science,\\
	C. V. Raman Avenue, Bangalore 560012, India.}
\emailAdd{justin@iisc.ac.in, srijankumar@iisc.ac.in}
\abstract{
We develop a method to  evaluate the 
partition function  and energy density of  a massive scalar
on a 2-sphere of radius $r$ 
and at finite temperature $\beta$  as power series in  $\frac{\beta}{r}$. 
Each term in the power series can be written in terms of polylogarithms. 
We use  this result to obtain the gap equation for the  large $N$, critical  $O(N)$ model  with a  quartic interaction
on $S^1\times S^2$ 
 in the  large radius expansion. 
 Solving the gap equation perturbatively 
 we obtain the leading finite size corrections to the expectation value of 
 stress tensor for the $O(N)$ vector  model on  $S^1\times S^2$. 
Applying the Euclidean inversion formula on the perturbative expansion of the thermal two point function we obtain 
the finite size corrections to the expectation value of the higher spin currents of the critical  $O(N)$ model. 
Finally we 
show that these finite size corrections of higher spin currents tend to that of the free theory at large spin
as seen earlier for the model on $S^1\times R^2$. }
\begin{document}
	\maketitle
	\flushbottom

		\section{Introduction}
	
	The critical $O(N)$ model of bosons
	with the quartic interaction in $d=3$ 
	 has been studied as an important model in quantum field theory in many contexts. 
	 It is known that this theory describes critical points of $O(N)$ Heisenberg ferromagnets and is conformal 
	 \cite{Brezin:1972se,Wilson:1973jj}. 
	 In the holographic context, the large $N$ limit of this model 
	  is conjectured to be the dual to the Vasiliev theory of higher spins in $AdS_4$
	 \cite{Klebanov:2002ja}, see \cite{Giombi:2012ms} for a review.

	 Since the model is relevant for condensed matter, there has been extensive studies of this model 
	 at finite temperature starting from \cite{Rosenstein:1989sg}. 
	 In \cite{Chubukov:1993aau,Sachdev:1993pr}, this theory was studied on $S^1\times R^2$ and 
	 it was shown that in the  large $N$ limit it admits a critical point and the value of stress tensor was obtained.
	 At the critical point, the model was shown to behave as a thermal CFT, that is the free energy and the 
	 energy density were proportional and temperature is the only scale in the theory. 
	 This model was 
	 generalised to complex scalars in the fundamental of $U(N)$ along with a chemical potential atfinite 
	 temperature in \cite{Filothodoros:2016txa,Filothodoros:2018pdj,Alvarez-Gaume:2019biu}. 
	 In this case too, the model admits a critical point. 
	 Recently  thermal  bootstrap methods have been applied to this model
	  and the one point function of arbitrary spin currents in this theory were obtained using the 
	  Euclidean inversion formula \cite{Iliesiu:2018fao,Petkou:2018ynm,David:2023uya,Karydas:2023ufs,David:2024naf,Barrat:2024fwq}. 
	  Furthermore it was shown that these one point functions approach that of the the free theory at 
	  large spin \cite{David:2023uya,David:2024naf}.

	 In spite of the extensive studies, there is no systematic  study of the model on $S^1\times S^2$. 
	 This is relevant  both as a field theory question as well as in holography.  
	 As a field theory, especially with regard to applications in condensed matter, systems always occur at finite 
	 size and it is interesting to study how the critical behaviour  and phase transitions are modified at finite size.  
	 In holography the behaviour of conformal field theories in the 
	 geometry $S^1\times S^2$ is  interesting.  Phase transitions in conformal field theories in this geometry are related 
	 to the Hawking-Page transitions  which provides insight to black hole physics \cite{Witten:1998zw}. 
	 
	 Recently there has been various efforts to investigate the general behaviour of thermal 
	 conformal field theories both in flat spatial geometries 
	  $S^1\times R^{d-1}$ and  on curved geometries such as $S^1\times S^{d-1}$ and 
	 $S^1\times H^{d-1}$ 
\cite{Shaghoulian:2016gol,Gobeil:2018fzy,Iliesiu:2018zlz,Karlsson:2021duj,Karlsson:2022osn,Kang:2022orq,Parisini:2022wkb,Luo:2022tqy,Dodelson:2023vrw,Benjamin:2023qsc,Marchetto:2023fcw,Parisini:2023nbd,Marchetto:2023xap,Allameh:2024qqp,Benjamin:2024kdg,Barrat:2024aoa,Buric:2024kxo}.  See \cite{Dowker:1983ci,Cardy:1991kr,Chang:1992fu} for  few examples of earlier work. 
These methods  rely on  the unbroken conformal 
symmetries and are usually applied to free theories or the gravitational dual. 
It is useful to have a  non-trivial  exactly solvable  thermal CFT which can be used to test the general behaviour derived 
using symmetries. 

Though the $O(N)$ vector model is not dual to Einstein gravity, it is exactly solvable at large $N$. 
However, its thermodynamics has not been studied when the theory is on a 2-sphere. 
In this paper we develop an approach with provides the expressions for its free energy, stress tensor and one point functions 
of higher spin currents as a power series in $\beta/r$, where $\beta$ is the temperature and $r$ is the radius of the sphere. 

We begin by deriving the partition function of the free 
massive scalar on the sphere. Surprisingly, we have not found this in the literature. 
Using similar steps  developed to write partition function of free fields on 
spheres in  terms of Harish-Chandra characters \cite{Anninos:2020hfj,David:2021wrw}, we arrive at the following expansion
\begin{align}\label{part fn order by orderi}
		\log Z\Big( \tilde m,\frac{\beta}{r}\Big)&=\frac{4\pi r^2}{\beta^2} \Bigg[\frac{\beta ^3 \tilde m^3+6 \beta  \tilde m \text{Li}_2\left(e^{-\tilde m \beta }\right)+6 \text{Li}_3\left(e^{-\tilde m \beta }\right)}{12 \pi }-\frac{\beta ^2 \left(\beta  \tilde m+2 \log \left(1-e^{-\tilde m \beta}\right)\right)}{96 \pi  r^2}\nonumber\\
		&\qquad\qquad\quad+\frac{\beta^4}{r^4}\frac{7\left(e^{\beta  \tilde m}+1\right)}{7680 \pi  \tilde m\beta \left(e^{\beta  \tilde m}-1\right)}
		+O\Big(\frac{\beta^6}{r^6}\Big)\Bigg].
\end{align}
where $\tilde m$ is the natural mass related to the  mass $m$ introduced in 
 the Lagrangian by a shift which depends on the curvature of the sphere. 
\begin{equation}
\tilde m^2 = m^2 - \frac{1}{4r^2} .
\end{equation}
Therefore $\tilde m^2$ measures the mass difference above the  mass induced  by the curvature coupling. 
 Observe that the  leading term in the expansion is the result of the partition function on $S^1\times R^2$. 
Though the higher order terms in $\beta/r$, appear to be rational functions of $e^{\tilde m\beta}$ and $\tilde m\beta$, each term can be written 
as sum of polylogarithms, just as seen for the thermodynamics of  massive theory on $S^1\times R^{d-1}$  \cite{Petkou:2021zhg}. 
The  all order expansion of the partition function  and the stress tensor in given in (\ref{allorderp}), (\ref{allordert}). 
As a check of this expression, we take the massless limit $\tilde m\rightarrow 0$, 
that is the limit in which the theory reduces to a conformally coupled 
scalar on $S^1\times R^2$ and we reproduce the result given in \cite{Benjamin:2023qsc}. 
This check involves using Borell summation to re-sum singular terms in the $\tilde m\rightarrow 0$ to obtain 
finite terms and is detailed in appendix \ref{C}.

Given the partition function of the massive scalar on $S^1\times S^2$, the partition function of the   $O(N)$ model at large 
$N$ is obtained by  extremising  with respect to the shifted  thermal mass $\tilde m$. This results in  the gap equation.
We have obtained the gap equation to all orders as an expansion in $\beta/r$ in (\ref{gapeq}).  
The  thermal mass can be obtained from the gap equation by solving it perturbatively in $\beta/r$ 
The  leading order expansion is given by 
\begin{eqnarray}
\tilde m\beta & = & 2 \log \frac{ \sqrt{5} + 1}{2} + \frac{1}{ 48 \csch^{-1} (2)}  \frac{\beta^2}{r^2} +  
\frac{55+64 \sqrt{5} \text{csch}^{-1}(2)}{230400 \text{csch}^{-1}(2)^3}  \frac{\beta^4}{r^4} + \cdots, \\ \nonumber
\csch^{-1} (2) &=& \log \frac{ \sqrt{5} + 1}{2} .
\end{eqnarray}
Note that the leading term for the thermal mass is the logarithm of the 
golden ratio. The expectation value of the stress tensor or the energy density
 can also be obtained perturbatively 
and is given by 
\begin{eqnarray}\label{E av}
\frac{\beta^3 }{4\pi r^2} \langle E \rangle_\beta  =  \frac{4\zeta(3)}{5\pi}-\frac{\beta^4}{r^4}\frac{1 }{288 \pi \sqrt{5}  \text{csch}^{-1}(2)}+O\big(\frac{\beta^6}{r^6}\big)
\end{eqnarray}
Again the leading term agrees with the result obtained in \cite{Chubukov:1993aau,Sachdev:1993pr} for the first time. 
Observe that there is no correction present at the  order $(\beta^2/r^2) $.  This implies that in the effective action approach of 
\cite{Benjamin:2023qsc,Allameh:2024qqp}  for  conformal field theories on $S^1 \times S^{d-1}$, the Wilson coefficient $c_1=0$.  It is consistent with the conjecture made  by \cite{Allameh:2024qqp}, that this coefficient obeys the 
bound $c_1 \geq 0$ for any conformal field theory.  

 We then  expand the Euclidean inversion formula perturbatively  to obtain finite size corrections to the 
 expectation value of the higher spin  currents 
 \begin{equation}
 {\cal O}[0, l] =  \vec\phi \cdot \partial_{\mu_1} \partial_ {\mu_2}  \ldots \partial_{ \mu_l} \vec \phi,
  \end{equation}
 where the indices $\mu_i$ are symmetric and traceless. 
For this we  first obtain the 2-point function of the massive scalar on $S^1\times S^2$ in position space. 
Here again we use similar steps followed in \cite{Anninos:2020hfj,David:2021wrw} to obtain a suitable integral form for the 
2-point function which admits a $\beta/r$ expansion. 
The propagator is given by following image sum 
\begin{equation}
g(\tau, \theta) = \sum_{n=-\infty}^\infty  \hat g(\tau + n \beta, \theta),
\end{equation}
where $r \theta$ is the geodesic distance on the $2$-sphere. 
In terms of the co-ordinates of 2 points on the sphere $(\theta_1, \phi_1)$ and $(\theta_2, \phi_2)$ the angle $\theta$ is 
 is given by 
\begin{eqnarray}
\cos \theta =  \Big[  \cos\theta_1 \cos\theta_2  + \sin\theta_1 \sin\theta_2 \cos ( \phi_1 - \phi_2) \Big],
\end{eqnarray}
and 
\begin{align}\label{massive corri}
		\hat g(\tau,\theta)=\frac{1}{4\pi \sqrt{2}\pi r } \left[ \frac{1}{(\cosh \frac{\tau}{r}-\cos\theta)^{1/2}}
		- \tilde m\int_\tau^\infty du\frac{u J_1( \tilde m\sqrt{u^2-\tau ^2})}{\sqrt{u^2-\tau ^2}(\cosh \frac{u}{r}-\cos \theta)^{1/2}}\right].
	\end{align}
Again we did not find such an expression for the propagator of a massive scalar on $S^1\times S^2$ in the literature. 
This form admits an expansion in inverse powers of the radius, which can be obtained by expanding the hyperbolic cosine in the
denominator.  We use this expansion to obtain a perturbative expansion of the higher spin one point functions. 
Demanding that the spin zero one point functions results in the gap equation, which agrees with that obtained from the partition function. 
The result  for the  infinite size corrections of one point functions  of  bi-linears with  spin $l >4$  
 is given by 
\begin{align}  \label{disc ansi}
		 a_{\cal O}^{(1)}[0,l] 
		= -	&\frac{ \Gamma( l +1) }{ 12 r^2 \sqrt{\pi} \Gamma( l + \frac{1}{2} ) }
		\Bigg[\sum _{n=0}^{l-4} \sum _{k=0}^{l-n-3} \frac{  \tilde m^k   (l-3)_{n+1} \text{Li}_{l-k-1}\left(e^{-\tilde m}\right)}{2^{l+n+2} k! n!}\nonumber\\
		&+\sum _{n=0}^{l-1} \frac{ (l-n)_{2n}    \tilde m^{l-n-1} \text{Li}_n\left(e^{-\tilde m}\right)}{2^{n+l+2} l! n! }\left(\frac{4 (l-1) l n}{(l+n-2) (l+n-1)}-1\right)\Bigg],\\
		{\rm for}\ l=  6, 8, \cdots .\nonumber
	\end{align}
	Here we have set $\beta =1$, we have to substitute the solution for the thermal mass obtained from the 
	gap equation. 
	The expressions for $l=4, 2, 0$ are given in section \ref{sec5}, as we mentioned earlier the 
	result for $l=0, 2$ agrees with that of the gap equation and the  stress tensor obtained directly 
	from the partition function. 
	For comparison,  here is the result for the expectation value of these bilinears at the leading order, that is when the 
	theory is on $S^1 \times R^2$ first obtained in \cite{Iliesiu:2018fao}, and also studied in \cite{David:2023uya,David:2024naf}. 
	\begin{align}
		&a_\mathcal{O}^{(0)}[0,l]
		=\frac{1}{2\pi \big(\frac{1}{2}\big)_l}\,\sum_{n=0}^l\frac{ \tilde m^{l-n} (l-n+1)_{2 n} }{2^{l+n} n! }\text{Li}_{n+1}
		\left(e^{-\tilde m+  \mu }\right)\\
		&l=2,4,6,\cdots.
	\end{align}
	The one point function is then given by 
	\begin{equation}
	a_{\cal O}[0, l ] =  a_{\cal O}^{(0)}[0,l]  + \frac{1}{r^2}  a_{\cal O}^{(1)}[0,l]  + \cdots. 
	\end{equation}
	and we have to substitute the thermal mass $\tilde m$ from the gap equation 
	including the $1/r^2$ correction.

One check of the perturbative expansion of the inversion formula and the fact that we are dealing with the 
thermal CFT is that it reduces to the stress tensor obtained from the partition function. 
We show that this leading correction also tends to that of the theory at the Gaussian fixed point given by setting 
$\tilde m =0$, 
on in the large $l$ limit.   This is in accordance with the earlier observations in \cite{David:2023uya,David:2024naf} 
where it was demonstrated that
higher spin one point functions of the large $N$ vector model $S^1\times R^2$ tend to their Gaussian fixed point values 
at large spin. 

The paper is organised as follows. 
In the next section \ref{sec2}, we derive the partition function of the massive scalar on $S^1\times S^2$. In section \ref{sec3}, 
we apply the result 
to the vector model at large $N$ and arrive at the gap equation and the energy density as an expansion in $\beta/r$. 
In section \ref{sec4}, we derive the expression given in (\ref{massive corri}) for the propagator of a massive scalar on $S^1\times S^2$ and 
set up and expansion in inverse powers of the radius. In section \ref{sec5}, we develop a perturbation theory for the higher spin 
one point function 
and obtain the leading finite size correction.  Section  \ref{sec6} contains our conclusions. 
	The appendix \ref{A} has the OPE of the propagator by taylor series expansion and shows the  agreement of the energy density with that from the partition function \eqref{E av}. The trick used to evaluate higher spin  one point functions  is being tested explicitly against a direct approach in appendix \ref{B}. The appendix \ref{C} contains the study of the massless limit, $\tilde m\to 0$,  of the partition function \eqref{part fn order by order}.

\section{Free  energy of a massive scalar on $S^1\times S^2$} \label{sec2}

We consider free massive  scalar field $\phi$ of mass $ m$ on $S^1_\beta\times S^2_r$, where $S^1_\beta$ is the thermal circle of length $\beta$ which is the inverse temperature and $r$ is the radius of the 2-dimensional sphere. 
The action is given by 
\begin{align}
	S=\frac{1}{2}\int_{S^1 \times S^2}  d^3 x \sqrt{g} \Big(\partial^\mu\phi\partial_\mu\phi+ m^2\phi^2\Big).
\end{align}
We take the metric to be 
\begin{eqnarray}
ds^2 = d\tau^2 + r^2 ( d\theta^2 + \sin^2 \theta d\phi^2) , \qquad \qquad \tau \sim \tau + \beta 
\end{eqnarray}
The partition function for the theory on $S^1\times S^2$ is given by the following path integral,
\begin{align}
	Z=\int_{S^1 \times S^2} D\phi e^{-\frac{1}{2} \int_0^\beta\int d^2 x\sqrt{g} (\partial^\mu\phi\partial_\mu\phi+ \tilde m^2\phi^2)}.
\end{align}
Performing the path integral we obtain 
\begin{align}
	\log Z=-\frac{1}{2}\sum_{n=-\infty }^\infty\sum_{l=0}^\infty(2l+1) \log\Big[\Big(\frac{2 n\pi}{\beta}\Big)^2+\frac{(l+\frac{1}{2})^2}{r^2}+ \tilde m^2\Big],
\end{align}
where $ \tilde m^2=m^2-\frac{1}{4r^2}$, the sum over $n$ represents the sum over Matsubara modes and $l$ denotes modes due to the spherical harmonics arising from the spherical harmonics on $S^2$. 
Note that as expected the  mass $m^2$ vanishes for a conformally coupled massless scalar.  
First we sum over the Matsubara modes using  the Sommerfeld-Watson transformation as
given in \cite{Klebanov:2011uf}.   This leaves us  with the sum over $l$ as given below,
\begin{eqnarray}\label{part fn sum}
	\log Z &=& - \frac{1}{2}\sum_{l=0}^{\infty} (2l+1) \Bigg(\beta\sqrt{\frac{(l+\frac{1}{2})^2}{r^2}+{\tilde m}^2}+2 \log \Big[1-e^{-\beta\sqrt{\frac{(l+\frac{1}{2})^2}{r^2}+ \tilde m^2} }\Big]\Bigg), \\ \nonumber
	&\equiv&  \log Z_1\big(-\frac{1}{2} \big)  +  \log Z_2.
\end{eqnarray}

Let us now develop the method to sum over the angular momentum $l$.  Focussing  on the  first term in (\ref{part fn sum}), 
we  rewrite it in the form as shown below,
\begin{eqnarray}
  \log Z_1 (\alpha) &=& 
	- \frac{1}{2}\sum_{l=0}^{\infty} (2l+1) 
	\Big({\frac{(l+\frac{1}{2})^2\beta^2}{r^2}+{\tilde m}^2\beta^2}\Big)^{-\alpha}, \\ \nonumber
	&=&-\frac{1}{2\Gamma(\alpha)} \sum_{l=0}^\infty (2l+1) \int_0^\infty {d\tau \tau^{\alpha-1}}e^{-[\frac{(l+\frac{1}{2})^2\beta^2}{r^2}+{\tilde  m}^2\beta^2]\tau}.
\end{eqnarray}
Here we have introduced the  dimensional regulator $\alpha$, in the end we will set $\alpha = -\frac{1}{2}$. 
This regulator is the same one used for performing the integral over the continuous momentum variable $k$ when the theory 
is placed on $S^1\times R^2$ which was used in \cite{Giombi:2019upv,David:2023uya,David:2024naf}. 
Next we  use of Hubbard-Stratanovich trick  to linearize the sum over $l$, this results in 
\begin{align}
  \log Z_1 (\alpha) 
	&=-\frac{1}{4 \sqrt{\pi}\Gamma(\alpha)} \int_0^\infty {d\tau \tau^{\alpha-1}} \sum_{l=0}^\infty (2l+1) e^{- \tilde m^2\beta^2\tau}\int_{-\infty}^{\infty}  \frac{du}{\sqrt{\tau}} e^{-\frac{u^2}{4\tau}+i\frac{u\beta}{r}(l+\frac{1}{2})}.
\end{align}
The sum over $l$ can be performed by introducing a small imaginary part to $u$ and then the integral over $\tau$ can
be carried out. This results in 
\begin{align}
  \log Z_1 (\alpha) 
	&=-\frac{r}{4 \sqrt{\pi}\Gamma(\alpha)\beta} \int_{-\infty}^\infty du\int_{0}^{\infty}{d\tau\tau^{\alpha-\frac{3}{2}}} e^{-\tilde m^2\beta^2\tau-\frac{u^2}{4\tau}} \frac{d}{du}\Big(\frac{1}{\sin \frac{u\beta}{2r}}\Big),\nonumber\\
	&=-\frac{r(\tilde m\beta)^{\frac{1}{2}-\alpha}}{2^{\alpha-\frac{1}{2}}\sqrt{\pi}\,\beta\Gamma(\alpha)} \int_{0}^\infty  u^{\alpha-\frac{1}{2}} K_{\alpha-\frac{1}{2}}( \tilde m\beta u ) \frac{d}{du}\Big(\frac{1}{\sin \frac{u\beta}{2r}}\Big).
\end{align}
To obtain the last line, we have used the fact that the integrand in $u$ is even. We can now integrate by parts to get
\begin{align}
	  \log Z_1 (\alpha) 
	&=-\frac{r( \tilde m\beta)^{\frac{3}{2}-\alpha}}{2^{\alpha-\frac{1}{2}}\sqrt{\pi}\beta\Gamma(\alpha)} \int_{0}^\infty  du  u^{\alpha-\frac{1}{2}} \frac{K_{\alpha-\frac{3}{2}}( \tilde m u \beta )}{\sin \frac{u\beta}{2r}}.
\end{align}
Now this integral cannot be evaluated easily in a closed form, however it is suitable for obtaining a $\beta/r$ expansion. 
This can be done  by expanding $\frac{1}{\sin \frac{u}{2r}}$ in the above integrand at large $r$. 
This leads to   the following power series  in $\beta/r$
\begin{eqnarray}
	  \log Z_1 (\alpha) 
	&=& -	\frac{r( \tilde m\beta)^{\frac{1}{2}-\alpha}}{2^{\alpha-\frac{1}{2}}\sqrt{\pi}\,\beta\Gamma(\alpha)} \sum_{p=0}^\infty \frac{2 (-1)^{p-1} \left(2^{2 p-1}-1\right) B_{2 p} }{(2 p)!} \\ \nonumber
	&& \qquad\qquad \times \int_{0}^\infty  du  u^{\alpha-\frac{1}{2}} {K_{\alpha-\frac{3}{2}}(\tilde m u \beta )} \times \Big(\frac{u\beta}{2r}\Big)^{2 p-1}. 
\end{eqnarray}
Finally substituting $\alpha=-\frac{1}{2}$ in the above expression, we obtain 
\begin{align} \label{z1sum}
  \log Z_1\Big(-\frac{1}{2} \Big) 
	&=-\sum_{p=0}^\infty\frac{  (-1)^p \left(4^p-2\right) B_{2 p} (\tilde m\beta)^{3-2 p} \big(\frac{\beta}{r}\big)^{2 p-2} \Gamma \left(p-\frac{3}{2}\right) \Gamma \left(p+\frac{1}{2}\right)}{4 \pi  (2 p)!}. 
\end{align}
It is clear that the expression obtained by substituting 
$\alpha=-\frac{1}{2}$ after performing the integral in (\ref{z1sum})  is an analytical continuation in $\alpha$. 
For the leading term in the $\beta/r$ series, this is the same analytical continuation used in \cite{Giombi:2019upv,David:2023uya,David:2024naf}. 
This completes the evaluation of the first term in (\ref{part fn sum}). 
Now we will evaluate the 2nd term from the partition function \eqref{part fn sum}. 
We carry out the following steps to linearise the sum over  $l$
\begin{align}
\log Z_2 &= 
	2 \sum_{l=0}^\infty \Big(l+\frac{1}{2}\Big) \sum_{n=1}^\infty \frac{e^{-n\beta\sqrt{\frac{(l+\frac{1}{2})^2+}{r^2}  \tilde m^2}}}{n}, \\
	\nonumber
	&=\frac{1}{\sqrt{\pi} } \sum_{l=0}^\infty \sum_{n=1}^\infty 
	\frac{1}{n}\Big(l+\frac   {1}{2}\Big) \int_0^\infty \frac{d\tau}{\tau^{3/2}}e^{-\tau n^2\beta^2[\frac{(l+\frac{1}{2})^2}{r^2}+ \tilde m^2]-\frac{1}{4\tau}}, \\ \nonumber
	&=\frac{1}{2\pi } \sum_{l=0}^\infty \sum_{n=1}^\infty 
	\frac{1}{n^2}\Big(l+\frac{1}{2}\Big) \int_0^\infty \frac{d\tau}{\tau^{2}}e^{-\tau n^2\beta^2  \tilde m^2-\frac{1}{4\tau}}\int_{-\infty}^\infty du e^{-\frac{u^2}{4\tau n^2}-i\frac{(l+\frac{1}{2})\beta u}{r}} . 
\end{align}
Performing the integral  over $\tau$ and the sum over  $l$,  we obtain 
\begin{align}
\log Z_2=
	\frac{  \tilde m r}{\pi} \sum_{n=1}^\infty \int_{-\infty}^\infty du \frac{K_1(\beta \tilde m \sqrt{n^2+u^2})}{\sqrt{n^2+u^2}} \frac{d}{du}\Big[\frac{1}{\sin \frac{u\beta}{2r}}\Big]. 
\end{align}
Finally,  integrating by parts
\begin{align}\label{2nd term 1}
\log Z_2 
	&=\frac{ \tilde m^2r\beta}{\pi} \sum_{n=1}^\infty \int_{-\infty}^\infty \frac{u du}{\sin \frac{\beta u}{2r}}   \frac{  K_2(\beta \tilde m \sqrt{n^2+u^2})}{n^2+u^2}.
\end{align}
Just as we did for the first term of the partition function, we 
expand the $\frac{1}{\sin \frac{\beta u}{2r}}$ in large $r$ and integrate term by term in each order in $\frac{1}{r}$ to obtain the finite sizer corrections  in all orders of $\frac{1}{r}$. This results in 
\begin{align} \label{z2fn}
&\log Z_2 = \\ \nonumber
&	\sum_{p=0}^\infty\left(\frac{\beta }{r}\right)^{2 p-2}\sum_{j=0}^{|p-\frac{3}{2}|-\frac{1}{2}} 	\frac{ (-1)^{p+1} \left(2^{2p-1}-1\right)  B_{2 p}   \left(\left| p-\frac{3}{2}\right| -j+\frac{1}{2}\right)_{2 j}  \text{Li}_{j-p+2}\left(e^{-\tilde m \beta }\right)}{2^{j+3p-2}j! \Gamma (p+1)(\tilde m\beta)^{j+p-1}}. 
\end{align}
Thus combining the 1st and 2nd term the partition function \eqref{part fn sum} using (\ref{z1sum}) and (\ref{z2fn}), we obtain the following 
$\beta/r$ expansion for the partition function
\begin{align} \label{allorderp}
	\log Z\Big(\tilde m,\frac{\beta}{r}\Big)=&\frac{4\pi r^2}{\beta^2}\sum_{p=0}^\infty \Big(\frac{\beta}{r}\Big)^{2 p}\frac{ (-1)^{p+1}}{4\pi} (2^{2p-1}-1)B_{2p}\times \\ \nonumber
	&\Bigg[ \frac{     \Gamma \left(p-\frac{3}{2}\right) \Gamma \left(p+\frac{1}{2}\right)}{2 \pi  (2 p)!( \tilde m \beta)^{2p-3} }
	+\sum_{j=0}^{|p-\frac{3}{2}|-\frac{1}{2}} 	\frac{    \left(\left| p-\frac{3}{2}\right| -j+\frac{1}{2}\right)_{2 j}  \text{Li}_{j-p+2}\left(e^{- \tilde m \beta }\right)}{2^{j+3p-2}j! \Gamma (p+1) ( \tilde m \beta)^{j+p-1}}\Bigg].
\end{align}
Expanding the above expression up to a few orders in $\frac{\beta}{r}$,
\begin{eqnarray}\label{part fn order by order}
		\log Z\Big( \tilde m,\frac{\beta}{r}\Big)&=&\frac{4\pi r^2}{\beta^2} \Bigg[\frac{\beta ^3 \tilde m^3+6 \beta  \tilde m \text{Li}_2\left(e^{-\tilde m \beta }\right)+6 \text{Li}_3\left(e^{-\tilde m \beta }\right)}{12 \pi }-\frac{\beta ^2 \left(\beta  \tilde m+2 \log \left(1-e^{-\tilde m \beta}\right)\right)}{96 \pi  r^2}\nonumber\\
		&& +\frac{\beta^4}{r^4}\frac{7\left(e^{\beta  \tilde m}+1\right)}{7680 \pi  \tilde m\beta \left(e^{\beta  \tilde m}-1\right)}
		+\frac{\beta^6}{r^6}\frac{31  \left(2 \beta  \tilde m e^{\beta  \tilde m}+e^{2 \beta  \tilde m}-1\right)}{258048 \pi  (\tilde m\beta)^3  \left(e^{\beta  \tilde m}-1\right)^2}+O\Big(\frac{\beta^8}{r^8}\Big)\Bigg].
\end{eqnarray}
Observe that the term on the first line is the partition function of a massive scalar on $S^1\times R^2$ \cite{Petkou:2021zhg,David:2024naf}. The subsequent terms are  the finite size corrections. 
The pre-factor $4\pi r^2$ can be identified as the spatial volume. 
The expansion of the partition function of a massive scalar on $S^1\times S^2$  given in 
 involved the steps in which the sum over $l$ was linearised by the 
Hubbard-Stratonovich transformation, this is where the we have been inspired by the methods developed in 
\cite{Anninos:2020hfj,David:2021wrw}. As  far as we are aware the expansion in (\ref{part fn order by order}) is new. 
The expectation value of the energy is given by 
\begin{align} \label{allordert}
	\langle E\rangle_\beta  &=-\partial_\beta \log Z\nonumber\\
	&=\frac{4\pi r^2}{\beta^3}\sum_{p=0}^\infty \Big(\frac{\beta}{r}\Big)^{2 p} \frac{(-1)^{p+1}}{4\pi} (2^{2p-1}-1)B_{2p}\times\nonumber\Bigg[- \frac{     \Gamma \left(p-\frac{3}{2}\right) \Gamma \left(p+\frac{1}{2}\right)}{2 \pi   (2 p)!(\tilde m \beta)^{2p-3} }+\\
	&\sum_{j=0}^{\left| p-\frac{3}{2}\right| -\frac{1}{2}}\frac{ \left(\left| p-\frac{3}{2}\right| +\frac{1}{2}-j\right)_{2 j}  }{2^{j+3p-2} (\tilde m\beta)^{j+p}j! \Gamma (p+1)} \Big(\beta  \tilde m \text{Li}_{j-p+1}(e^{-\tilde m \beta })+(j-p+1) \text{Li}_{j-p+2}(e^{-\tilde m \beta })\Big)\Bigg].
\end{align}
Expanding the energy to a few orders in $\beta/r$, we obtain 
\begin{eqnarray}
	\langle E \rangle_\beta
	&=&\frac{4\pi r^2}{\beta^3}\Bigg[\frac{-\tilde m^2\beta^2 \left(\beta  \tilde m+6 \log \left(1-e^{-\tilde m\beta}\right)\right)+12 \beta  \tilde m \text{Li}_2\left(e^{-\tilde m \beta }\right)+12 \text{Li}_3\left(e^{-\tilde m \beta }\right)}{12 \pi }\nonumber\\
	&&\qquad\qquad+\frac{\beta^2}{r^2}\frac{\tilde m\beta\left(e^{\beta  \tilde m}+1\right)}{96 \pi   \left(e^{\beta  \tilde m}-1\right)}
	-\frac{\beta^4}{r^4}\frac{7 \left(e^{2 \beta  \tilde m}-2 \beta  \tilde m e^{\beta  \tilde m}-1\right)}{7680 \pi  \tilde m\beta \left(e^{\beta  \tilde m}-1\right)^2}\nonumber\\
	&&+\frac{\beta^6}{r^6}
	\frac{31 \left(e^{\beta  \tilde m} \left(2 \beta ^2 \tilde m^2+2 \beta  \tilde m+1\right)+e^{2 \beta  m} \left(2 \beta ^2 \tilde m^2-2 \beta  \tilde m+1\right)-e^{3 \beta  \tilde m}-1\right)}{258048 \pi  (\tilde m\beta)^3  \left(e^{\beta  \tilde m}-1\right)^3} \nonumber \\ 
	&& \qquad \qquad \qquad +O\Big(\frac{\beta^8}{r^8}\Big)\Bigg].
\end{eqnarray}
Here again, the leading term agrees precisely with the expectation value of the energy of a massive boson on $S^1\times R^2$
in a volume $4\pi r^2$ \cite{Petkou:2021zhg,David:2024naf}. 
The remaining terms are the sub-leading  finite size  corrections. 
The time component of the stress tensor can be read out from the relation 
\begin{eqnarray}
T_{00} = \frac{\langle E \rangle_\beta}{4\pi r^2}.
\end{eqnarray}

\subsection*{$ \tilde m\to 0 $ limit of the partition function}

As a check of the result for the partition function, we take the limit $ \tilde m \rightarrow 0$. 
In this limit the result should reduce to the  conformally coupled scalar. 
The first term in (\ref{part fn sum}) reduces to 
\begin{eqnarray}
\lim_{\tilde m\rightarrow 0} \log Z_1\big( -\frac{1}{2} \big)  &=& - \frac{\beta}{4 r} \sum_{l =0}^\infty ( 2l +1)^2 , \\ \nonumber
&=&  \frac{3 \beta}{4 r}  \zeta (-2)  = 0 
\end{eqnarray}
For the  second term, we take the limit $\tilde m\rightarrow 0$ in the integral in (\ref{2nd term 1}), this results in 
\begin{eqnarray} \label{mzerolim}
\lim_{\tilde m\rightarrow 0}  \log Z \big(\tilde m, \frac{\beta}{r} \big) &=&  \lim_{\tilde m\rightarrow 0}  \log Z_2  , \\ \nonumber
&=& 
	\frac{2r}{\pi\beta} \sum_{n=1}^\infty \int_{-\infty}^\infty \frac{u du}{\sin \frac{\beta u}{2r}}   \frac{ 1}{(n^2+u^2)^2},  \\ \nonumber
	&=&\frac{1}{2} \sum _{n=1}^{\infty } \frac{\coth \frac{\beta  n}{2 r}\ \text{csch}\frac{\beta  n}{2 r}}{ n }.
\end{eqnarray}
This  coincides with the result for the massless conformally coupled scalar in \cite{Chang:1992fu,Benjamin:2023qsc}\footnote{See equation (C.19) of \cite{Benjamin:2023qsc}.} . We also can  take the $\tilde m\to 0$ limit in \eqref{allorderp} and present the massless free partition function on $S^1\times S^2$ as  systematic order by order corrections to the  partition function on $S^1\times R^2$ as was given in equation (C.22) of 
\cite{Benjamin:2023qsc}. The naive $\tilde{ m}\to 0$ limit in \eqref{allorderp} gives rise to a divergent series where each term of the series diverges at $\tilde m\to 0$, but we have shown in appendix \ref{C} that these infinite number of diverging terms combines to give a finite value to this series at this limit by using the method of Borel sum.

\section{O(N) model on $S^1 \times S^2 $} \label{sec3}

Let us consider the 
 model of $N$ scalars $\phi$ with quartic interaction on $S^1\times S^2$, the $O(N) $ invariant action for the model is given by,
\begin{align}\label{O(N) action}
	S= \frac{1}{2} \int d^3x\sqrt{g}  [\partial^\mu\phi_i\partial_\mu\phi_i+\frac{\lambda}{N} (\phi_i\phi_i)^2], 
\end{align}
where
	$i=1,\cdots,N$ and $\lambda $ characterises the interaction strength. The partition function on $S^1\times S^2 $ is given by,
	\begin{align}
	\tilde	Z=\int_{S^1 \times  S^2} {\cal D}\phi e^{- \frac{1}{2} \int_0^\beta\int d\tau d^2 x \sqrt{g} 
	[\partial^\mu\phi_i\partial_\mu\phi_i+\frac{\lambda}{N} (\phi_i\phi_i)^2]} .
	\end{align}
We apply the  Hubbard-Stratanovich transformation by introducing an auxiliary field $\zeta$ to linearise the quartic interaction term in the action
\begin{align}
\tilde	Z=\int {\cal D}\zeta\int_{S^1 \times S^2 } {\cal D}\phi e^{-\frac{1}{2} \int_0^\beta\int d\tau d^2 x [\partial^\mu\phi_i\partial_\mu\phi_i+\frac{\zeta^2 N}{4\lambda}+i\zeta\phi_i\phi
		_i]} .
\end{align}
We have re-written the quartic interaction  by introducing the path integral in the auxiliary field $\zeta$ so that the resulting action turns out to be quadratic in the field $\phi_i$. Now separating the zero mode of the auxiliary field from its non-zero modes in the following manner
\begin{align}
	\zeta=\zeta_0+\tilde \zeta, 
\end{align}
where $\zeta_0$ denotes the zero mode of the auxiliary field $\zeta$ and the non-zero modes of $\zeta$ are contained in $\tilde \zeta$. Now the zero mode can be treated as a mass of the scalar field in the theory with an integral over $\zeta_0$. And the 
	\begin{align}\label{int zero mode}
	\tilde Z[\zeta_0] &=  \int d\zeta_0  \exp\left[  -\beta 4\pi r^2 N \Big( \frac{ \zeta_0^2}{4\lambda}  - \frac{1}{4\pi r^2\beta} \log Z ( \zeta_0,\frac{\beta}{r})
	\Big)
	\right],
\end{align}
where
\begin{align}
	\log Z( \zeta_0,\frac{\beta}{r})=-\frac{1}{2}\sum_{n=-\infty }^\infty\sum_{l=0}^\infty(2l+1) \log\Big[\Big(\frac{2 n\pi}{\beta}\Big)^2+\frac{(l+\frac{1}{2})^2}{r^2}+ \tilde m^2\Big], 
\end{align}
with 
\begin{equation} \label{relationmzeta}
\tilde m^2 = i \zeta_0 - \frac{1}{4r^2} . 
\end{equation}
From   \eqref{part fn order by order}, we can expand this partition function  as an expansion in $\beta/r$,  
 and has the following expression
	\begin{align}\label{log Z}
		\log Z\Big(\tilde m,\frac{\beta}{r}\Big)=&\frac{4\pi r^2}{\beta^2}\sum_{p=0}^\infty \Big(\frac{\beta}{r}\Big)^{2 p}\frac{ (-1)^{p+1}}{4\pi} (2^{2p-1}-1)B_{2p}\times\nonumber\\
		&\Bigg[ \frac{     \Gamma \left(p-\frac{3}{2}\right) \Gamma \left(p+\frac{1}{2}\right)}{2 \pi  (2 p)!( \tilde m \beta)^{2p-3} }
		+\sum_{j=0}^{|p-\frac{3}{2}|-\frac{1}{2}} 	\frac{    \left(\left| p-\frac{3}{2}\right| -j+\frac{1}{2}\right)_{2 j}  \text{Li}_{j-p+2}\left(e^{- \tilde m \beta }\right)}{2^{j+3p-2}j! \Gamma (p+1) ( \tilde m \beta)^{j+p-1}}\Bigg].
	\end{align}
At large $N$, the leading contribution to the partition function $\tilde Z$  is obtained by performing the saddle point integration 
over $\zeta_0$ in \eqref{int zero mode}. 
  The resulting saddle point equation in terms of $ \tilde m$ is given by 
\begin{align}
	\partial_{\tilde m} \log \tilde Z( \tilde m)=0.  
\end{align}
At this point we would to mention that we could have also minimised the partition function with respect to $\zeta_0$, the  resulting 
equation would be same due to the relation (\ref{relationmzeta}). However the equation organises in a more convenient form 
in terms of $\tilde m$.  
At strong coupling, $\lambda\to \infty$ this equation reduces to 
\begin{align}
	\partial_{ \tilde m} \log Z( \tilde m,\frac{\beta}{r})=0. 
\end{align}
Substituting 
the expression for $\log \tilde Z( \tilde m,\frac{\beta}{r})$ given in \eqref{log Z} in the saddle point condition  we obtain
\begin{align} \label{gapeq}
&\sum_{p=0}^\infty	(-1)^p \Big(\frac{\beta}{r}\Big)^{2p} \left(4^p-2\right) B_{2 p}\Big[\frac{  \Gamma \left(p-\frac{1}{2}\right) \Gamma \left(p+\frac{1}{2}\right)}{\pi  (2 p)!}+\nonumber\\
&\sum_{j=0}^{\left| p-\frac{3}{2}\right| -\frac{1}{2}}\frac{  \left(\left| p-\frac{3}{2}\right| +\frac{1}{2}-j\right)_{2 j} }{2^{j+3p-2}j!p!(\beta \tilde m)^{j-p+2}}  \left(\beta  \tilde m \text{Li}_{j-p+1}(e^{-\tilde m \beta })+(j+p-1) \text{Li}_{j-p+2}(e^{-\tilde m \beta })\right)\Big]=0. 
\end{align}
Expanding this equation to a  few orders in $\frac{\beta}{r}$,  the  gap equation  is given by 
\begin{eqnarray} \label{gapeqorderby}
	&& \frac{r^2}{\beta^2} \tilde m\beta\left(\beta \tilde  m+2 \log (1-e^{- \tilde m \beta})\right)+\frac{e^{\beta  \tilde m}+1}{24(1- e^{\beta  \tilde m})}
	\\ \nonumber
	&& \qquad\qquad -\frac{\beta^2}{r^2}\frac{7 \left(2 \beta  \tilde m e^{\beta  \tilde m}+e^{2 \beta  \tilde m}-1\right)}{1920 (\tilde m\beta)^2  \left(e^{\beta  \tilde m}-1\right)^2}+O(\frac{\beta^4}{r^4})=0. 
\end{eqnarray}

\subsection*{ Perturbative solution of the gap equation}
To obtain the perturbative solution of the gap equation as an expansion in $\beta/r$, we expand the thermal mass
\begin{align} \label{mexpan}
	\tilde m= \frac{1}{\beta}\Big(c+\frac{c_1\beta}{r}+\frac{c_2\beta^2}{r^2}+\frac{c_3\beta^3}{r^3}+\frac{c_4\beta^4}{r^4}+O\big(\frac{\beta^5}{r^5}\big)\Big)
\end{align}
The coefficients $c,c_1,c_2,c_3,c_4,\cdots$ are determined by the condition that $ m$ satisfies the gap equation order 
by order in $\beta/r$. 
At the leading order that is matching terms at $O(r^2/\beta^2)$, we obtain 
\begin{eqnarray} \label{orderl}
	&&2 \beta  c^2+4 \beta  c \log \left(1-e^{-c}\right)=0 \qquad\qquad\qquad\qquad\qquad\quad  :O(r^2/\beta^2) \nonumber\\
	&&c=2\log \frac{1+\sqrt{5}}{2}
\end{eqnarray}
At the subsequent orders of matching terms at $O(r/\beta)$, and $O(\beta/r)^0$, we obtain
\begin{eqnarray} \label{orderll}
	&& 4 \beta  c c_1+4 \beta  \left(\frac{c c_1}{e^c-1}+c_1 \log \left(1-e^{-c}\right)\right)=0,   \qquad\qquad\qquad  :O(r/\beta) 
	\nonumber \\ 
	&& c_1=0
\end{eqnarray}
and 
\begin{eqnarray}
	&&\frac{\beta  \left(24 c_1^2\left(e^{2 c}-e^c c-1\right) +\left(e^c-1\right) \left(e^c (48 c c_2-1)-1\right)\right)}{12 \left(e^c-1\right)^2}+4 \beta  c_2 \log \left(1-e^{-c}\right)=0,  \nonumber \\
	&& \qquad\qquad\qquad\qquad\qquad\qquad\qquad\qquad \qquad\qquad\qquad\qquad\qquad : O(  ( \beta/r)^0 ) 
\end{eqnarray}
Using the solutions $c$ and $c_1=0$ from (\ref{orderl}) and (\ref{orderll}) we obtain 
\begin{eqnarray}
c_2=\frac{1}{48 \text{csch}^{-1}2}.
\end{eqnarray}
Proceeding similarly we obtain 
\begin{align}
	c_3=0, \qquad  c_4=\frac{55+64 \sqrt{5} \text{csch}^{-1}(2)}{230400 \text{csch}^{-1}(2)^3}.
\end{align}
Substituting  these coefficients in the expansion  (\ref{mexpan}), we obtain the following expansion  for the thermal 
mass
\begin{align}\label{gap soln}
	 \tilde m=\frac{1}{\beta}\Big(2 \log\frac{1+\sqrt{5}}{2}+\frac{\beta^2}{r^2}\frac{1}{48 \text{csch}^{-1}2} +\frac{\beta^4}{r^4}\frac{55+64 \sqrt{5} \text{csch}^{-1}2}{230400 (\text{csch}^{-1} 2)^3}+O\big(\frac{\beta^6}{r^6}\big)\Big).
\end{align}
Using this expansion 
in the expression for partition function given in (\ref{log Z}), we obtain 
\begin{align} \label{logZs}
	\log Z\Big(\tilde m,\frac{\beta}{r}\Big)=\frac{4\pi r^2}{\beta^2}\Big[\frac{2 \zeta (3)}{5 \pi }+\frac{\beta^4}{r^4}\frac{1}{576 \sqrt{5} \pi   \text{csch}^{-1}(2)}+O\big(\frac{\beta^6}{r^6}\big)\Big] .
\end{align}
The expectation value of the energy is given by 
\begin{eqnarray}\label{energy}
	\langle E\rangle_\beta&=&-\partial_\beta	\log Z(\tilde m,\frac{\beta}{r})\nonumber\\
	&=&\frac{4\pi r^2}{\beta^3}\Big[\frac{4\zeta(3)}{5\pi}-\frac{\beta^4}{r^4}\frac{1 }{288 \pi \sqrt{5}  \text{csch}^{-1}(2)}+O\big(\frac{\beta^6}{r^6}\big)\Big].
\end{eqnarray}

It is important to point out, that the expansion is a systematic   perturbative expansion in $\beta/r$ and can in principle be carried 
out to all orders.  The general form for the expectation value of the stress tensor of a CFT on a sphere is given by 
\begin{equation}
T_{00} = \frac{1}{\beta^3} f( \frac{\beta}{r}) .
\end{equation}
Therefore this method allows us to construct the non-trivial function $f( \frac{\beta}{r})$ as a perturbative expansion. 
Another observation is the behaviour of pressure which is given by 
\begin{eqnarray}
P &=& \frac{1}{\beta } \frac{\partial Z}{\partial V}  =  \frac{1}{\beta }   \frac{\partial Z}{4\pi \partial r^2 } , \\ \nonumber
&=&  \frac{2 \zeta (3)}{5 \pi  \beta ^3}	-\frac{1}{576  \sqrt{5}  \pi  \text{csch}^{-1}(2)}\frac{\beta}{r^4}+ \cdots
\end{eqnarray}
Note that the correction to the pressure is negative, which is expected when a  bosonic system is confined. 

It is illustrative to compare the result in (\ref{logZs}) for the partition function, with that of the theory at the Gaussian fixed point, 
For that we need to take $\tilde m \rightarrow 0$. 
This can be obtained by a systematic expansion of the partition function given in (\ref{mzerolim}), which
is shown in appendix \ref{C}.
\begin{eqnarray}\label{Duffin}
\lim_{\tilde m\rightarrow 0} \log Z\big( \tilde m , \frac{\beta}{r} \big)  = 
\frac{ 2 r^2 \zeta(3) }{\beta^2}  - \frac{1}{12} \log ( \frac{\beta}{r} )   - \frac{\log 2}{12} - \zeta'(-1) + 
\frac{ 7}{ 11520} \frac{\beta^2}{r^2} + \cdots.
\end{eqnarray}
Note that presence of the $\log T $ term which arises due to contribution of the Matsubara zero mode for which the theory reduces to 
a free boson in $2d$.  Comparing with (\ref{logZs}), we see the departure of the critical strong coupling CFT, 
 from the free CFT result, 
the leading term is different by the famous  factor $\frac{4}{5}$ \cite{Chubukov:1993aau,Sachdev:1993pr}. 
However the sub-leading terms are   different  numerically as well as 
qualitatively  due to the presence of the $\log T $ term. 

\subsection*{Density of states}

The density of states of the $O(N)$ model at the Wilson-Fisher fixed point at large  $N$ was discussed in 
\cite{Sachdev:2023try}. Using earlier results of Wilson and Fisher \cite{Wilson:1971dc} as well as the result of the
$T^2$ behaviour of the entropy density, the energy density was shown  to change as follows
\begin{eqnarray}
\rho (E) && \sim r, \qquad\qquad rE\ll1, \\ \nonumber
\rho(E) && \sim r \exp{ (rE)^{3/2 } }, \qquad \qquad rE\gg1
\end{eqnarray}
In \cite{Sachdev:2023try}, this behaviour was compared with the SYK model and 
it was suggested that this transition in the density of states may be due to the presence of `black hole' like states. 
Since we have finite size  corrections to the free energy it is interesting to obtain the 
 finite size corrections to density fo states 
at $rE\gg1$ 
\footnote{This question was raised by Subir Sachdev after the first version of this paper appeared on the arXiv.}. 
To proceed with this we use the relation of the density of states to the partition function 
 \begin{eqnarray}
 \rho(E) =\frac{1}{2\pi} \int_{\delta -i\infty}^{ \delta+i \infty}  d\beta Z(\beta) e^{\beta E}, 
 \end{eqnarray}
where the $\delta$ is chosen so that  the poles are to the left of the contour. 
From (\ref{logZs}), we have the following result for the partition function
\begin{eqnarray}  \label{defAB}
\log Z(\beta, r) = 
\frac{A}{\hat\beta^2}  + B \hat \beta^2 + O(\hat \beta^4) , \\ \nonumber
 \hat \beta = \frac{\beta}{r}, \qquad A = \frac{ 8 \zeta(3)}{5} , \qquad B =\frac{1}{144\sqrt{5} {\rm csch}^{-1} ( 2) }. 
 \end{eqnarray}
 After a  re-scaling of the  integration variable, we  can write the Laplace transform as 
 \begin{eqnarray} \label{saddleint}
 \rho(E) &=& \frac{r}{2\pi} \int_{a-i\infty}^{ a+\infty} 
  d\hat\beta \exp \left( \hat \beta \hat E +  \frac{A}{\hat\beta^2}  + B \hat \beta^2 \right) , \\
 \nonumber
 \hat E &=& rE. 
 \end{eqnarray}
In the limit 
\begin{eqnarray} \label{lim}
 \hat E= r E \gg1
 \end{eqnarray}
 we can perform the integral in the saddle point approximation. 
 We are interested in 
keeping track of the dependence of the sub-leading term, that is  all the contribution 
 proportional to $B$. 
 The saddle point  occurs at the  solution of the following equation 
 \begin{eqnarray}
 \hat \beta^3 - \frac{2A}{\hat E} + \frac{2B}{\hat E} \hat\beta^4 =0.
 \end{eqnarray}
 We  solve this equation perturbatively in the limit (\ref{lim}) .  Let us define
 \begin{equation}
 \epsilon  ^3 = \frac{2A}{\hat E} . 
 \end{equation}
 The solution to the saddle point  is given by 
 \begin{eqnarray}
 \hat \beta_0 &=& \epsilon( 1 - \frac{B}{3A} \epsilon^4 + \cdots), \\ \nonumber
 &=& \Big(\frac{2A}{\hat E}\Big)^{1/3} \left[  1 - \frac{B}{3A} \Big(\frac{2A}{\hat E}\Big)^{4/3}  + O\big(\hat E^{-2} \big) \right] .
 \end{eqnarray}
 The value of the exponent in (\ref{saddleint})  at the saddle point is given by 
  \begin{eqnarray}
 \left.  \hat \beta\hat  E + \frac{A}{\hat \beta^2}  + B\hat \beta^2 \right|_{\hat\beta_0} =
  \frac{3}{2}   (2A)^{1/3} \hat E^{\frac{2}{3} }   +   B\Big(  \frac{2A}{\hat E}\Big)^{2/3} 
+ O\big( \hat E^{-2} \big) .
  \end{eqnarray}
 The second derivative of the exponent in (\ref{saddleint}) at the saddle point is given by 
 \begin{eqnarray}
   \left.  \frac{6A}{\hat\beta^4} + 2 B\right |_{\hat \beta_0}
  = 6A \Big( \frac{\hat E }{2 A}\Big  )^{4/3}  \Bigg[
  1+ \frac{5B}{3A} \Big( \frac{2A}{\hat E}   \Big) ^{\frac{4}{3} } + O\big(\hat E^{-2}\big) \Bigg].
  \end{eqnarray}
  We need to chose the direction of the contour at the saddle point to 
 to be in the $y$ direction so that the second derivative  is negative. 
 Using all these inputs, we find the density of states at $rE\gg1$ to be  given by 
  \begin{eqnarray}
 \rho(E) &=& \frac{r}{ \sqrt{6 \pi} } \Big(  \frac{2A }{\hat E } \Big)^{\frac{2}{3}} 
{ \left[ 1 +B \Big(  \frac{2A}{\hat E} \Big)^{2/3}  + \Big(  - \frac{5}{6} \frac{B}{A} + \frac{B^2}{2} \Big)
 \Big(  \frac{2A}{\hat E} \Big)^{4/3} + O(\hat E^{-2})  \right]  }  \nonumber \\
  && \qquad\qquad\qquad \times \exp\left( {  \frac{3}{2}  ( 2A)^{1/3} \hat E^{2/3}} \right)  .
 \end{eqnarray}
 Thus the finite size corrections to the partition function modify the behaviour of the  pre-factor to the exponent in the 
 density of states at high energies. It will be interesting to obtain the behaviour of the density of states 
 as a function of $rE$ for all regimes and see how the exponential behaviour of the density of states 
 at high energies transitions to 
the order one behaviour at low energies.

	\section{Two point function of a massive scalar on $S^1\times S^2$} \label{sec4}
	
	The two point function of a massive scalar field in position space on $S^1\times S^2$ can be written in terms of 
	the Matsubara sum on $S^1$ and the spherical harmonics on $S^2$.  This is given by 
	\begin{eqnarray}
	g(\tau, \theta) =  \frac{1}{4\pi r^2 \beta } \sum_{n=-\infty}^\infty \sum_{l =0}^\infty 
	\frac{ (2l +1) e^{ \frac{ 2\pi i  n \tau} {\beta} } P_l ( \cos\theta) }{ \big(  \frac{ 2\pi n }{\beta} \big)^2 +
	 \frac{1}{r^2} ( l + \frac{1}{2} )^2 + \tilde m^2 }. 
	\end{eqnarray}
	The normalization by the volume $4\pi r^2 \beta$ 
	not only ensures the right dimension of the correlator, but also reproduces the normalization 
	on $S^1\times R^2$ used earlier in \cite{Iliesiu:2018fao,David:2024naf} in the $r\rightarrow \infty $ limit. 
	The length $r\theta$ measures the geodesic length on the sphere and is given by 
	\begin{eqnarray}
	cos \theta = \cos\theta_1 \cos\theta_2 + \sin\theta_1\sin\theta_2 \cos(\phi_1 - \phi_2) . 
	\end{eqnarray}
	where $(\theta_1, \phi_1)$ and $( \theta_2, \phi_2)$ are the 2 points are the sphere. 
	Performing a Poisson re-summation over $n$, we can write the 2-point function as a sum over images
	\begin{eqnarray}\label{imagsum}
	g(\tau, \theta) = \sum_{n=-\infty}^\infty \hat g ( \tau + n \beta, \theta) 
	\end{eqnarray}
	where 
	\begin{eqnarray}
		\hat g(\tau,\theta)=\frac{1}{8\pi^2 r^2  }\int_{-\infty}^\infty  d\omega e^{i\omega\tau}\sum_{l=0}^\infty \frac{(2l+1)P_l(\cos \theta)}{\omega^2+\frac{1}{r^2}(l+\frac{1}{2})^2+  \tilde m^2} . 
		\end{eqnarray}
		We  use  Schwinger parametrisation  to exponentiate the denominator resulting in 
		\begin{eqnarray}
		\hat g(\tau,\theta)
		=\frac{1}{8\pi^2 r^2  }\sum_{l=0}^\infty (2l+1)P_l(\cos \theta) \int_{-\infty}^\infty 
		d\omega e^{i\omega\tau} \int_0^\infty dze^{-z[\omega^2+\frac{1}{r^2}(l+\frac{1}{2})^2+ {m}^2]} 
	\end{eqnarray}
	We then linearise the dependence on the angular momentum by the Hubbard-Stratonovich transformation 
	\begin{eqnarray} \label{hubbard}
		\hat g(\tau,\theta)
		&=& \frac{1}{16 \pi^{5/2} r^2 }\sum_{l=0}^\infty (2l+1)P_l(\cos \theta) \int_{-\infty}^\infty
		d\omega e^{i\omega\tau}\Big[   \\ \nonumber
		&& \qquad\qquad \qquad  \int_0^\infty \frac{dz}{\sqrt{z}}e^{-z[\omega^2+\tilde {m}^2]} 
		\int_{-\infty }^{\infty } du e^{-\frac{u^2}{4 z}+\frac{i (l+\frac{1}{2}) u}{r}}  \Big].
	\end{eqnarray}
	Performing the  integral over $\omega$ we obtain
	\begin{align}
		\hat g(\tau,\theta)
		&=\frac{1}{16 \pi^{2} r^2 }\sum_{l=0}^\infty (2l+1)P_l(\cos \theta)  \int_0^\infty \frac{dz}{{z}}e^{-z{m}^2} 
		\int_{-\infty }^{\infty } du e^{-\frac{u^2+\tau^2}{4 z}+\frac{i (l+\frac{1}{2}) u}{r}} .
	\end{align}
	The integral over $z$   results in modified Bessel function of 2nd kind, 
	\begin{align} \label{besselstep}
		\hat g(\tau,\theta)
		&=\frac{1}{8\pi^{2} r^2}\sum_{l=0}^\infty (2l+1)P_l(\cos \theta)  
		\int_{-\infty }^{\infty } du e^{\frac{i (l+\frac{1}{2}) u}{r}}   K_0 ( \tilde m \sqrt{u^2+\tau ^2}).
	\end{align}
	The sum over the angular momentum can be carried out using the identity
	\begin{eqnarray}
	\sum_{l=0}^\infty (2l+1)P_l(\cos \theta)  e^{\frac{i (l+\frac{1}{2}) u}{r}}  = \frac{\sqrt{2} r }{i} \frac{\partial}{\partial u} 
	\frac{1}{\sqrt{ \cos\frac{u}{r} - \cos\theta } }.
	\end{eqnarray}
	Here, we assume $u$ has a small positive imaginary component so that the sum is convergent. 
	Substituting, this identity in (\ref{besselstep}), we get
	\begin{align}
		\hat g(\tau,\theta)
	&=\frac{1 }{i4 \sqrt{2}\pi^{2}r }
	\int_{-\infty }^{\infty } du    K_0 ( \tilde m \sqrt{u^2+\tau ^2})	\frac{d}{du}\Big[\Big(\cos \frac{u}{r}-\cos\theta \Big)^{-1/2}\Big].
	\end{align}
	The integral runs  along the contour $C$ from $-\infty$ to $\infty$ slightly above the real axis, since $u$ has a small 
	imaginary part. The integrand has a branch cut which runs along the positive 
	imaginary axis from $i\tau$ to $\infty$ as 
	shown in the figure \ref{fig 1}.  We then replace $u= i {u'} $, where $
	{u'}$ is real. Then the integral runs along the contour $C'$ as shown in \ref{fig 2}, the cut is now along the real axis. 
	Since the integrand is analytic  in the right half of the $u'$ plane 
	and dies off sufficiently fast at infinity, we can deform the contour to that 
	shown as $C''$ in the figure \ref{fig 2}, which encircles the cut on the right of the $u'$ plane. 
	\begin{figure}[h]
		\centering
		\begin{tikzpicture}[thick,scale=0.85]
			\draw [decorate,decoration=snake] (0,-0.7) -- (0,-4);
			\draw [decorate,decoration=snake] (0,0.7) -- (0,4);
			
			\draw
			[
			postaction={decorate,decoration={markings ,
					mark=at position 0.20 with {\arrow[red,line width=0.5mm]{>};}}}
			][red, thick] (-4,0.2)--(0,0.2);
			\draw[red, thick] (0,0.2)--(4,0.2);
			\draw (0.4,1) node{$\mathbf{i \tau}$};
			\draw (0.4,-1) node{$\mathbf{-i \tau}$};
			\draw[gray, thick] (0,0) -- (0,4);
			\draw[gray, thick] (0,0) -- (0,-4);

			\draw[gray, thick] (0,0) -- (4,0);
			\draw[gray, thick] (0,0) -- (-4,0);
			
			\draw (2.3,3.3) node{$\mathbf{u}$};
			\draw [red,thick](-2,0.5) node{$\mathbf{C}$};
			
		\end{tikzpicture}
		\caption{The original integral along $C$ in $u$-plane} \label{fig 1}
		\qquad
		
		\centering
		\begin{tikzpicture}[thick,scale=0.85]
			
			\draw
			[
			postaction={decorate,decoration={markings ,
					mark=at position 0.80 with {\arrow[red,line width=0.5mm]{>};}}}
			][red, thick] (.2,-4) -- (.2,4);
			\draw [decorate,decoration=snake] (-0.7,0) -- (-4,0);
			\draw [decorate,decoration=snake] (0.7,0) -- (4,0);
			\draw
			[
			postaction={decorate,decoration={markings ,
					mark=at position 0.20 with {\arrow[red,line width=0.5mm]{<};}}}
			][blue, thick] (0.7,0.2)--(4,0.2);
			\draw
			[blue, thick] (0.7,-0.2)--(4,-0.2);
			\draw (1,0.4) node{$\mathbf{\tau}$};
			\draw (-1,0.4) node{$\mathbf{-\tau}$};
			\draw[gray, thick] (0,0) -- (0,4);
			\draw[gray, thick] (0,0) -- (0,-4);
			\draw[gray, thick] (0,0) -- (4,0);
			\draw[gray, thick] (-4,0) -- (0,0);
			
			\draw (2.7,3.3) node{$\mathbf{u'=iu}$};
			\draw [postaction={decorate,decoration={markings ,
					mark=at position 0.55 with {\arrow[black,line
						width=0.5mm]{};}}}](0.7,.2) arc[start angle=90, end angle=270,
			radius=0.2cm];
			\draw [red,thick](0.6,3) node{$\mathbf{C'}$};
			\draw [blue,thick](3,0.6) node{$\mathbf{C''}$};
		\end{tikzpicture}
		\caption{Using $u'=iu$ the integral is along $C'$ and finally the integral along $C'$ is -ve of the integral along $C''.$} \label{fig 2}
	\end{figure}
	
	The integral then picks up the discontinuity across the cut on the real line. To evaluate the discontinuity, we have the  identity 
	\begin{eqnarray}
	\pi i J_0(z) = K_0(-iz) - K_0(i z)
	\end{eqnarray}
	Using this, we find that the two point function is given by 
	\begin{eqnarray}
	\hat g(\tau,\theta) = - \frac{1 }{4\sqrt{2}\pi^{}r } \int_{\tau}^\infty du J_0( \tilde m \sqrt{u^2 - \tau^2}) \frac{d}{du}\Big[\Big(\cosh \frac{u}{r}-\cos\theta \Big)^{-1/2}\Big].
	\end{eqnarray}
	Finally integrating by parts, we obtain 
	\begin{align}\label{massive corr}
		\hat g(\tau,\theta)=-\frac{1}{4 \sqrt{2} \pi r} \left[\frac{J_0(\tilde m\sqrt{u^2-\tau^2})}{(\cosh \frac{u}{r}-\cos\theta)^{1/2}}\bigg|_{\tau}^\infty+\tilde m\int_\tau^\infty du\frac{u J_1( \tilde m\sqrt{u^2-\tau ^2})}{\sqrt{u^2-\tau ^2}(\cosh \frac{u}{r}-\cos \theta)^{1/2}}\right]. 
	\end{align}
	Using the fact that the boundary term at $\infty$ vanishes we can write the correlator as 
		\begin{align}\label{massivecorr1}
		\hat g(\tau,\theta)=\frac{1}{4 \sqrt{2} \pi r} \left[\frac{1}{(\cosh \frac{\tau}{r}-\cos\theta)^{1/2}}
		- \tilde m\int_\tau^\infty du\frac{u J_1( \tilde m\sqrt{u^2-\tau ^2})}{\sqrt{u^2-\tau ^2}(\cosh \frac{u}{r}-\cos \theta)^{1/2}}\right]. 
	\end{align}
	We can use this expression for the $\hat g(\tau, \theta)$ into (\ref{imagsum}) to obtain the 2-point function of a massive 
	scalar on $S^1\times R^2$. 
	As far as we are aware, we have not found this 
	representation of the 2-point function of a massive scalar in literature. 
	One simple test  of the formula in (\ref{massive corr}) is to take the $\tilde m\rightarrow 0$  limit, this results in 
	\begin{equation}\label{2ptzeromass}
	\lim_{m\rightarrow 0}  \hat g(\tau,\theta) = \frac{1}{4\sqrt{2} \pi r} \frac{1}{(\cosh \frac{\tau}{r}-\cos\theta)^{1/2}}.
	\end{equation}
	We see that this coincides with the propagator of the conformally coupled massless boson  found on $R\times S^2$ 
	in \cite{Pufu:2013eda} \footnote{See equation (32) of \cite{Pufu:2013eda}}, 
	which was obtained using a conformal transformation from $R^3$. 
	
	It is important to mention that the method by which we obtained the expression for the 2 point function in 
	(\ref{massivecorr1}) relied on the the Hubbard-Stratonovich trick in performing the sum over the angular
	momenta in (\ref{hubbard}) as well as the contour deformation shown in figures . 
	These methods  are an adaptation of the steps used to obtain partition function of free fields on spheres 
	in terms of integrals over Harish-Chandra characters \cite{Anninos:2020hfj,David:2021wrw}. 

\subsection*{Large radius expansion}

We can now proceed to expand the two point function in (\ref{massivecorr1})  as a power series in $\beta/r$. 
We  expand the $\cosh$  terms of its power series at the origin
We also  define 
\begin{equation}\label{chord}
r^2 ( 1- \cos \theta) = \tilde \theta^2 , 
\end{equation}
$\tilde \theta$ has dimensions and is equal to the chordal distance on the sphere.  We keep $\tilde \theta$ finite 
in the large radius expansion. 
This distance coincides with the 
distance on $R^2$, when the radius of the sphere is infinite. 
	The 1st term from of (\ref{massivecorr1})  can be expanded as, 
	\begin{eqnarray}\label{1st term}
		\frac{1}{(\cosh \frac{\tau}{r}-\cos\theta)^{1/2}}
		&=&\frac{\sqrt{2}r}{(\tau^2+\tilde\theta^2)^{1/2}} \Big[1-\frac{\tau ^4}{24 (\tilde \theta ^2+\tau ^2)
			r^2} \\ \nonumber 
			&&\quad \qquad+\frac{1}{r^4}\Big(\frac{\tau ^8}{384 (\tilde \theta ^2+\tau
			^2)^2}-\frac{\tau ^6}{720 (\tilde \theta ^2+\tau
			^2)}\Big)+\cdots\Big].
	\end{eqnarray}
	Performing the expansion of the $\cosh\frac{u}{r}$ in the integrand of the 
	2nd term of \eqref{massive corr}, we obtain
	{\small \begin{eqnarray}\label{2nd term}
		&&\tilde  m\int_\tau^\infty du\frac{u J_1( m\sqrt{u^2-\tau ^2})}{\sqrt{u^2-\tau ^2}(\cosh \frac{u}{r}-\cos \theta)^{1/2}}
		=\tilde m 
		\sqrt{2} r \int_\tau^\infty du\frac{u J_1(\tilde  m\sqrt{u^2-\tau ^2})}{\sqrt{u^2-\tau ^2}\sqrt{u^2+\tilde \theta^2}}   \\ \nonumber
		&& \qquad\qquad\qquad\qquad\qquad\qquad\qquad  \times  \left[1-\frac{u ^4}{24 (\tilde \theta ^2+u ^2)
			r^2}+\frac{1}{r^4}\Big(\frac{u ^8}{384 (\tilde \theta ^2+u
			^2)^2}-\frac{u ^6}{720 (\tilde \theta ^2+u
			^2)}\Big)+\cdots\right].
	\end{eqnarray}}
	We can now perform the integral term by term. After the integration, it can be seen that upon 
	adding \eqref{1st term} and \eqref{2nd term}  there is a term by term cancellation of the expansion 
	in  \eqref{1st term}.  
	This results in the following expansion of the two point function
	{\small \begin{align} \label{larg2ptexp}
	&\hat g(\tau,\theta)\nonumber\\
	=&-\frac{\tilde m }{4\pi } 
	\Bigg[-\frac{e^{-\tilde m \sqrt{\tilde \theta ^2+\tau ^2}}}{ \tilde m \sqrt{\tilde \theta ^2+\tau ^2}}+\frac{e^{- \tilde m \sqrt{\tilde \theta ^2+\tau ^2}} \left(-\left(\tilde \theta ^2+\tau ^2\right)^{3/2}+\tilde \theta ^4  \tilde m^2 \sqrt{\tilde \theta ^2+\tau ^2}-\tilde m \left(2 \tilde \theta ^2 \tau ^2+\tilde \theta ^4\right)\right)}{24 \tilde m^2 r^2 \left(\tilde \theta ^2+\tau ^2\right)^{3/2}}\nonumber\\
		&+\frac{1}{r^4} \frac{e^{- \tilde m \sqrt{\tilde \theta ^2+\tau ^2}} }{5760  \tilde m^4 \left(\tilde \theta ^2+\tau ^2\right)^{5/2}}\Big(-21 \left(\tilde \theta ^2+\tau ^2\right)^{5/2}-36 \tilde \theta ^2  \tilde m^2 \left(\tilde \theta ^2+\tau ^2\right)^{5/2}-5\tilde  \theta ^8  \tilde m^5 \left(\tilde \theta ^2+\tau ^2\right)\nonumber\\
		&- \tilde m^3 \left(80 \tilde \theta ^6 \tau ^2+66 \theta ^4 \tau ^4+29\tilde  \theta ^8\right)+\tilde \theta ^6  \tilde m^4 \left(37 \tilde \theta ^2+52 \tau ^2\right) \sqrt{\tilde \theta ^2+\tau ^2}-21  \tilde m \left(\tilde \theta ^2+\tau ^2\right)^3\Big)+O\Big(\frac{1}{r^6}\Big)\Bigg]  .
	\end{align} }
	Observe that the leading term coincides with the $2$ point function on $R^3$ once  we have 
	identified $\tilde \theta$ to be the distance on the spatial $R^2$. 
	Using this expansion in the image sum  (\ref{imagsum}), we obtain the large radius expansion of the propagator 
	of a massive scalar on $S^1\times R^2$. 
	We could have chosen to perform the expansion so that the geodesic distance $r\theta$ appears in the 
	combination $\tau^2 + r^2\theta^2$ instead of $\tau^2 + \tilde\theta^2$. 
	However we have seen that it is the chordal distance $\tilde\theta$ defined in (\ref{chord}) which appears 
	naturally. This is also the natural distance that the massless scalar propagator in (\ref{2ptzeromass}) is expressed. 
	We will subsequently see  in section (\ref{sec5}), 
	that it is the expansion in terms of the chordal distance that the finite size corrections 
	of the gap equation and that of the stress tensor obtained from the partition function 
	agree with that obtained form the Euclidean inversion formula.

			\section{Thermal one point function using OPE inversion formula} \label{sec5}
			
			In this section we will use the  large radius expansion of the 2-point function 
			massive scalar on $S^1 \times S^2$
			given in (\ref{larg2ptexp}) to obtain the finite size corrections to one point functions 
			of higher spin currents in the $O(N)$ model at its strong coupling critical point. 
			To do this we would need to first formulate the OPE expansion of the two point function
			in terms of one point functions on $S^1\times S^2$. 
			By the symmetries on $S^1\times S^2$, we can write the expectation values of 
			 one point functions of symmetric 
			traceless tensors  of rank $l$  as  \cite{Iliesiu:2018fao}, 
			\begin{eqnarray}
			\langle {\cal O}^{\mu_1, \mu_2, \cdots \mu_l }[n,l] \rangle_{S^1\times S^2}
			= \frac{ b_{ {\cal O}[n,l]  }} {\beta^{\Delta_{{\cal O} [n,l]   }}}  f_{{\cal O}[n,l] }\big( \frac{\beta}{r} \big) 
			\big( e^{\mu_1} \cdots e^{\mu_l}  - {\rm Trace} \big) .
			\end{eqnarray}	
			Here $f_{{\cal O}[n,l] }(\frac{\beta}{r})$ is  a non-trivial function that depends on the CFT and the operator 
			${\cal O}[l]$. $e^{\mu}$ is the unit vector  in the Euclidean time direction. 
			This result is essentially due to the fact that no tensor with indices valued in the $S^2$ direction 
			is invariant under the isometries of $S^2$. 
			The next step is to perform an OPE expansion of the two point function in which the symmetric 
			tensors appear. 
			For the $O(N)$ model, they appear in the scalar two point function which  admits the following expansion
			\begin{align} \label{opexpan}
				g(\tau,\vec{x})=\sum_{\mathcal{O} \in \phi \times \phi} 
				\frac{a_\mathcal{O}}{\beta^\Delta} C_l^{(\nu)}(\eta) |x|^{\Delta_{\cal O} -2\Delta_\phi}.
			\end{align}
			where  the length
			$x$ and $\eta$ and $\nu$  are given by 
			\begin{eqnarray}
			|x|^2 = \tau^2 + \tilde \theta ^2 , \qquad \eta = \frac{\tau}{ |x| }, \qquad \nu = \frac{d-2}{2} = \frac{1}{2}. 
			\end{eqnarray}
			$C_l^{(\nu)}(\eta)$ are Gengenbauer polynomials  and 
			\begin{eqnarray}
			a_{ {\cal O}} = \frac{f_{\phi\phi {\cal O} }  b_{\cal O} }{ c_{{\cal O}}}  \frac{l!}{2^l ( \nu)_l }, 
			\qquad\qquad\qquad   (a)_n = \frac{\Gamma( a+n)}{\Gamma(a) } 
			\end{eqnarray}
			Here $c_{{\cal O}}$ is the normalization of the two point function of  the  exchanged operator
			${\cal O}$ and $f_{\phi\phi{\cal O}}$ is the OPE coefficient in the corresponding 3-point function. 
			
			Note that we have chosen to perform  the OPE expansion in terms of the chordal length. 
			This is because the large radius expansion of the $2$-point function naturally organises in terms of the 
			chordal length. 
			Furthermore,  
			the inversion formula  which was developed  earlier \cite{Iliesiu:2018fao,Petkou:2018ynm,David:2023uya,David:2024naf}  assumed a specific  behaviour 
			of the two point function. We will see that the same behaviour holds 
			term by term  in the large radius expansion in terms of the chordal length. 
			We will also see that the one point function as well as the stress tensor derived from the 
			inversion formula obtained using the chordal length agrees with that of the partition function.

			The inversion formula is summarised as follows.
			Let us define  the complex variables 
			\begin{align} \label{defcomplex}
				z=\tau+i \tilde \theta = \tilde r w, \qquad\bar z=\tau-i\tilde \theta = \tilde r w^{-1}. 
			\end{align}
			 Observe from  the expansion (\ref{larg2ptexp}), 
			  the two point function can be written in terms of these variables. Once the two point function 
			 satisfies the condition that it has the branch cut structure pointed out in \cite{Iliesiu:2018fao} 
			 in the $w$-plane together with certain fall of properties, the inversion formula can be applied. 
			 The one point functions are given by 
				\begin{eqnarray} 
				a_\mathcal{O}[n,l]=-\hat a( \Delta_{} , l)|_{{\rm Res\ at\ }\Delta=\Delta_\mathcal{O}} &=& - [\hat a_{\rm disc} ( \Delta_{}, l) + \theta( l_0 -l) \hat a_{\rm arc} ( \Delta_{} ,l)]|_{{\rm Res\ at\ }\Delta=\Delta_\mathcal{O}} , \\ \nonumber
				& \equiv& \left. a_{{\cal O}}[n, l]\right|_{\rm disc}  +\left. a_{ {\cal O}}[n, l ]\right|_{\rm arc}. 
			\end{eqnarray}
			Here 
			$ a_{\rm disc}(\Delta,l) $ is related to the two point function by the following integral in $z, \bar z$. 
			 by,
			\begin{align}\label{a disc}
				&\hat a_{\rm disc}(\Delta,l)=K_l \int_0^1 \frac{d\bar z}{\bar z}\int_1^{1/\bar z}\frac{dz}{z} (z \bar z)^{\Delta_\phi-\frac{\Delta}{2}-\nu}(z-\bar z)^{2\nu} F_l\bigg(\sqrt{\frac{z}{\bar z}}\bigg) {\rm disc} [{g}(z,\bar z)], \\
				&\text{with,}\quad \nonumber
				K_l=\frac{\Gamma(l+1)\Gamma(\nu)}{4\pi \Gamma(l+\nu)},
				\qquad F_l(w) = w^{l+ 2\nu } {}_2 F_1( l + 2\nu , \nu , l + \nu +1, w^2) , 
			\end{align}
			and
			\begin{align} \label{Disc}
				{\rm disc}[ g(z, \bar z) ] = \frac{1}{i} \big(  g( z +i \epsilon , \bar z) - g( z-i\epsilon, \bar z) \big) .
			\end{align}
			The contribution due to the contour at infinity in the $w$ arises from 
			\begin{eqnarray}\label{arc}
				\hat a_{\rm arcs} (\Delta_{},  l)  &=&  2 K_l   \int_0^1 \frac{d\tilde r}{\tilde r^{\Delta_{}  +1 - 2\Delta_\phi} }  \times \nonumber \\ 
				&& \oint \frac{dw}{i w} \lim_{|w| \rightarrow \infty} 
				\left[ \Big( \frac{ w - w^{-1} }{i} \Big)^{2\nu} 
				F_l(w^{-1}) e^{i\pi\nu}  {g}( \tilde r, w) 
				\right].
			\end{eqnarray}
			
			We proceed to apply the inversion formula term by term to the expansion in  (\ref{larg2ptexp}). 
			For the leading term, the analysis has been already been carried out  in \cite{Iliesiu:2018fao}. 
			We consider the first subleading term in the large radius expansion and apply the 
			inversion formula to it.  
\begin{align}
	g^{(1)} (\tau,\theta)
	=-\sum_{n=-\infty}^\infty
	\frac{e^{-\tilde m \sqrt{\tilde \theta ^2+(\tau +n) ^2}}}{96\pi r^2\tilde m} \Big[&-1+\frac{\tilde \theta^4\tilde{ m}^2}{\tilde \theta^2+(\tau+n)^2}-\frac{2\tilde m \tilde \theta^2}{\sqrt{(\tau+n^2)+\tilde \theta^2}}
	+\frac{\tilde m \tilde \theta^4}{(\tilde \theta^2+(\tau+n)^2)^{\frac{3}{2}}}\Big].
\end{align}
We have set $\beta =1$ for convenience. 
Our approach is to  relate the subleading terms to the leading term by certain 
			operations and then apply the inversion formula.  Using the  definitions in (\ref{defcomplex}), 
			we have
\begin{align}\label{g^1}
	g^{(1)} (z,\bar z)
	&=-\frac{1}{96 \pi  \tilde m r^2} \sum_{n=-\infty}^\infty e^{-\tilde m\sqrt{(n-z)(n-\bar z)}}\Big[\frac{(z-\bar z)^4\tilde{ m}^2}{(2i)^4[(n-z)(n-\bar z)]}\nonumber\\
	&\qquad\qquad\qquad +\frac{\tilde m (z-\bar z)^4}{(2i)^4[(n-z)(n-\bar z)]^{\frac{3}{2}}} -\frac{2\tilde m (z-\bar z)^2}{(2i)^2\sqrt{(n-z)(n-\bar z)}}
	-1\Big],\\
	&\equiv g_1(z, \bar z)+
	g_2(z, \bar z)+g_3(z, \bar z)+g_4(z, \bar z).	\nonumber		\end{align}
	\subsection*{Disc contribution}
			We will be applying the inversion formula \eqref{a disc} separately on each of the terms inside parenthesis in equation \eqref{g^1} as it is described below.
			The following formula will be useful to evaluate the discontinuity of the four terms in the correlator $g^{(1)}(\tau,\theta)$ across its branch cut,
			\begin{align} \label{discontinuity}
			{\rm disc}\Big[\frac{e^{-\tilde m \sqrt{(m-z)(m-\bar z)}}}{[(m-z)(m-\bar z)]^\alpha}\Big]=	\frac{2 \sin \left(\pi  \alpha +\tilde m \sqrt{(z-m) (m-\bar z)}\right)}{((z-m) (m-\bar z))^{\alpha }}.
			\end{align}
			First let us examine the first term in the equation \eqref{g^1} and we rewrite it as follows
						\begin{align}\label{int 1st term}
				g_1(z,\bar z)=&-\frac{\tilde m}{96 \pi  r^2}
				\sum_{n=-\infty}^\infty \frac{e^{-\tilde m\sqrt{(n-z)(n-\bar z)}}(z-\bar z)^4}{(2i)^4[(n-z)(n-\bar z)]}\nonumber,\\
				&=\frac{\tilde m}{96 \pi  r^2} \sum_{m=-\infty}^\infty \int_{\infty}^{\tilde m} d\tilde m\frac{e^{-\tilde m\sqrt{(m-z)(m-\bar z)}}(z-\bar z)^4}{(2i)^4[(m-z)(m-\bar z)]^{1/2}}.
			\end{align}
			To apply the inversion formula \eqref{a disc} on the integrand from the above expression \eqref{int 1st term}, we evaluate   the disc contribution which is given by \footnote{It is possible to directly use the result for the 
			discontinuity in (\ref{discontinuity}) in the first line of (\ref{int 1st term}) with $\alpha=1$ and proceed with the 
			inversion formula. The final result  involves hypergeometric functions of the type ${}_1F_2$. 
			This is not as illuminating as the approach we have chosen to demonstrate here. In the appendix \ref{B} we have verified both these approaches yields  the same final result. }
			\begin{align}
I=K_l	\int_0^1 \frac{d\bar z}{\bar z}\int_m^{{\rm max}(m,\frac{1}{\bar z})}\frac{d z}{ z}
			\frac{(z-\bar z)^{5}  F_l\left(\sqrt{\frac{\bar z}{z}}\right)  2 \cos \left(\tilde m \sqrt{(z-m) (m-\bar z)}\right) }{(2i)^4(z\bar z)^{\Delta/2} ((z-m) (m-\bar z))^{1/2 }}.
			\end{align}
			To evaluate this integral , we  follow the same steps as developed in \cite{Iliesiu:2018fao}. 
			We first use the following coordinate transformation,
			\begin{align}
				 \bar z=z z',\qquad \qquad{\rm and}\qquad z=m z'.
			\end{align}
			to obtain,
			\begin{align}
			I=	-K_l\int_0^1 d\bar z\int_1^{{\rm max}(1,\frac{1}{m\sqrt{\bar z}})} dz\frac{m^6 z^6 (\bar z-1)^5 F_l\left(\sqrt{\bar z}\right)   \cos \left(m \tilde m\sqrt{(1-z) (z \bar z-1)}\right) }{8(m^2z^2\bar z)^{\frac{\Delta}{2}+1} \sqrt{(1-z) (z \bar z-1)}}.
			\end{align}
			We then expand  the integrand  around $\bar z=0$. 
			It is only the leading term that contains the 
			poles in $\Delta$. For the leading term, the integral over $z$ and $\bar z$ decouples. 
			Performing the  integral over  $\bar z$   on this term. 
			 we are left with the following $z$ integral,
			\begin{align}
			I^{(0)}=K_l	\int_1^\infty dz\frac{m^4 z^4 \left(m^2 z^2\right)^{-\frac{\Delta }{2}} \cos \left(m \tilde m \sqrt{z-1}\right)}{4 \sqrt{z-1} (-\Delta +l+1)}.
			\end{align}
			Now with the use of the change of variable given below
			\begin{align}
				z'^2=z-1,
			\end{align}
this integral is evaluated  in terms of  modified Bessel function of 2nd kind. 
 The   residue of the resulting expression at $\Delta=l+1$ is given by 
			\begin{align} \label{residue}
			-I^{(0)}|_{{\rm Res\ at\ } \Delta=l+1}&=	-K_l\Big[\int_0^{\infty } \frac{\left(m \left(z^2+1\right)\right)^{4-\Delta } \cos (m \tilde m z)}{2 (-\Delta +l+1)} \, dz\Big]_{{\rm Res\ at\ }\Delta=l+1},\\
			&=K_l\frac{\sqrt{\pi } 2^{\frac{5}{2}-l} m^{-l} (m \tilde m)^{l-\frac{1}{2}} K_{l-\frac{7}{2}}(m \tilde m)}{\tilde m^3 \Gamma (l-3)}.
			\end{align}
	The modified Bessel function of 2nd kind with half integer index  can be written  terms of a finite sum  as follows
	\begin{align}
	K_{l-\frac{7}{2}}(m \tilde m)=	\sum _{n=0}^{\left| l-\frac{7}{2}\right| -\frac{1}{2}} \frac{\sqrt{\pi } e^{-m \tilde m}  \left(\left| l-\frac{7}{2}\right|-n +\frac{1}{2}\right)_{2 n}}{2^{n+\frac{1}{2}}n! (m \tilde m)^{n+\frac{1}{2}}}.
	\end{align}
	Substituting this identity into (\ref{residue}), we get 
	\begin{align}
		-I^{(0)}|_{{\rm Res\ at\ } \Delta=l+1}=K_l\sum _{n=0}^{\left| l-\frac{7}{2}\right| -\frac{1}{2}}\frac{\pi   e^{-m \tilde m} \left(\left| l-\frac{7}{2}\right|-n +\frac{1}{2}\right)_{2 n} (m \tilde m)^{l-n-1}}{m^l2^{l+n-2}\tilde m^3 n! \Gamma (l-3)}.
	\end{align}
	At this stage we perform the integral over $\tilde m$ as shown below
	\begin{align}
	-\int_{\infty}^{\tilde m}d{\tilde m}I^{(0)}|_{{\rm Res\ at\ } \Delta=l+1}&=K_l	\sum _{n=0}^{\left| l-\frac{7}{2}\right| -\frac{1}{2}}\int_\infty^{\tilde m} d \tilde m	\frac{\pi   e^{-m \tilde m} \left(\left| l-\frac{7}{2}\right|-n +\frac{1}{2}\right)_{2 n} (m \tilde m)^{l-n-1}}{m^l2^{l+n-2}\tilde m^3 n! \Gamma (l-3)}\nonumber,\\
	&=-K_l\sum _{n=0}^{\left| l-\frac{7}{2}\right| -\frac{1}{2}} \frac{\pi \left(\left| l-\frac{7}{2}\right|-n +\frac{1}{2}\right)_{2 n} \Gamma (l-n-3,m \tilde m)}{m^{l-2}2^{l+n-2}n! \Gamma (l-3)}.
	\end{align}
	Now reinstating all the constant factors and sum over $m$ from \eqref{int 1st term} the contribution from  $g_1(z,\bar z)$ to the disc part of the inversion formula is given by 
	\begin{align}\label{a_1}
		\left. a_{\mathcal{O},1}^{(1)}[0,l]\right|_{\rm disc} =-	\frac{K_l \tilde{ m}}{3 r^2}
		\sum _{n=0}^{\left| l-\frac{7}{2}\right| -\frac{1}{2}} \sum_{m=1}^\infty \frac{ \left(-n+\left| l-\frac{7}{2}\right| +\frac{1}{2}\right)_{2 n} \Gamma (l-n-3,m \tilde m)}{m^{l-2}2^{l+n+2}n! \Gamma (l-3)}.
	\end{align}
	To clarify the notation $a_{\mathcal{O},1}^{(1)}[0,l]$: the $(1)$ in the superscript   denotes that we are evaluating the first  subleading correction in $\frac{1}{r^2}$ to the one point functions and the $1$ in the subscript stands for the contribution due to $g_1(z,\bar z)$ given in \eqref{g^1}.
Now using similar techniques, we apply the inversion formula to evaluate the contribution 
from  $g_2(z,\bar z)$, $g_3(z,\bar z)$ and $g_4(z,\bar z)$ to the first subleading correction to the thermal one point function. 

The contribution from  $g_2(z,\bar z)$ to the first subleading correction in $\frac{1}{r^2}$ to the one point function  is given by \footnote{To evaluate $a_{{\cal O},2 }^{(1)}[0,l]$, we use similar trick as used in \eqref{g^1}, $g_{2}(z,\bar z)$ is expressed as the following
\begin{align}
	g_2(z,\bar z)=-\frac{1}{96\pi r^2(2i)^4}\int_\infty^{\tilde{m}}dy\int_\infty^y dx \frac{e^{-x\sqrt{(n-z)(n-\bar z)}}}{\sqrt{(n-z)(n-\bar z)}}
	\end{align}    
Now we apply the OPE inversion formula \eqref{a disc} on the integrand and perform the integrals at the end to obtain the result \eqref{a_2}. \\For computing $a_{{\cal O},3}^{(1)}[0,l]$ and $a_{{\cal O},4}^{(1)}[0,l]$ we do not require such tricks we can apply the inversion formula \eqref{a disc} directly on $g_3(z,\bar z)$ and $g_4(z,\bar z)$ respectively.  }
\begin{align}\label{a_2}
	\left. a_{\mathcal{O},2}^{(1)}[0,l]\right|_{\rm disc} =\frac{K_l}{3 r^2}
	\sum _{n=0}^{\left| l-\frac{7}{2}\right| -\frac{1}{2}}\sum_{m=1}^\infty &\frac{   \left(\frac{1}{2}-n+\left| l-\frac{7}{2}\right| \right)_{2 n} }{m^{l-1}2^{l+n+2}n! \Gamma (l-3)}\times\nonumber\\
	&\big[m \tilde m \Gamma (l-n-3,m \tilde m)-\Gamma (l-n-2,m \tilde m)\big].
\end{align}
Combining  \eqref{a_1} and \eqref{a_2} we get, 
\begin{align}
\left. \big(a_{\mathcal{O},1}^{(1)}[0,l]+a_{\mathcal{O},2}^{(1)}[0,l]\big)\right|_{\rm disc} =-\frac{K_l}{3r^2}
	\sum_{n=0}^{|l-\frac{7}{2}|-\frac{1}{2}}\sum_{m=1}^\infty\frac{ \left(-n+\left| l-\frac{7}{2}\right| +\frac{1}{2}\right)_{2 n} \Gamma (l-n-2,m \tilde m)}{m^{l-1}2^{n+l+2} n! \Gamma (l-3)}.
\end{align}
Now we use 
\begin{align}
	\Gamma(l-n-2,m \tilde m)=\sum _{k=0}^{l-n-3} \frac{e^{-m \tilde m} (m \tilde m)^k \Gamma (l-n-2)}{k!}.
\end{align}
to obtain
\begin{align}\label{1+2}
	&\left. \big(a_{\mathcal{O},1}^{(1)}[0,l]+a_{\mathcal{O},2}^{(1)}[0,l]\big) \right|_{\rm Disc} \nonumber\\
	&=-\frac{K_l}{ 3r^2}
	\sum _{n=0}^{\left| l-\frac{7}{2}\right| -\frac{1}{2}} \sum _{k=0}^{l-n-3} \frac{  \tilde m^k \Gamma (l-n-2) \left({|  l-\frac{7}{2}| }-n+\frac{1}{2}\right)_{2 n} \text{Li}_{l-k-1}\left(e^{-\tilde m}\right)}{2^{l+n+2} k! n! \Gamma (l-3)}.
	\end{align}
	The contribution due to $g_3(z,\bar z)$ and $g_4{(z,\bar z)}$ are respectively obtained to be
\begin{align}\label{3}
	\left. a_{\mathcal{O},3}^{(1)}[0,l] \right|_{\rm Disc}=-\frac{K_l}{ 3r^2}\sum_{n=0}^{|l-\frac{3}{2}|-\frac{1}{2}}\frac{   \tilde m^{l-n-2} \text{Li}_{n+1}\left(e^{-\tilde m}\right) \left({|  l-\frac{3}{2}| }-n+\frac{1}{2}\right)_{2 n}}{2^{l+n+1} n! \Gamma (l-1)}.
\end{align}
and finally 
\begin{align}\label{4}
	\left. a_{\mathcal{O},4}^{(1)}[0,l]\right|_{\rm disc}
	=\frac{K_l}{3 r^2}\sum_{n=0}^{|l-\frac{1}{2}|-\frac{1}{2}}\frac{   \tilde m^{l-n-1}  \left({|  l-\frac{1}{2}| }-n+\frac{1}{2}\right)_{2 n}}{2^{l+n+2} n! \Gamma (l+1)}\text{Li}_n\left(e^{-\tilde m}\right).
\end{align}

	Summing the contributions \eqref{1+2}, \eqref{3} and \eqref{4}
	to the leading correction to the  thermal one point function of spin-$l$ bi-linears 
	we obtain
	\begin{align}  \label{disc ans}
		\left. a_{\cal O}^{(1)}[0,l] \right|_{\rm disc}
		= -	&\frac{ K_l}{ 3r^2}\Bigg[\sum _{n=0}^{l-4} \sum _{k=0}^{l-n-3} \frac{  \tilde m^k   (l-3)_{n+1} \text{Li}_{l-k-1}\left(e^{-\tilde m}\right)}{2^{l+n+2} k! n!}\nonumber\\
		&+\sum _{n=0}^{l-1} \frac{ (l-n)_{2n}    \tilde m^{l-n-1} \text{Li}_n\left(e^{-\tilde m}\right)}{2^{n+l+2} l! n! }\left(\frac{4 (l-1) l n}{(l+n-2) (l+n-1)}-1\right)\Bigg],\\
		{\rm for}\ l\ge 4.\nonumber
	\end{align}
	Note that the 1st term is due to $\big(a_{\mathcal{O},1}^{(1)}[0,l]+a_{\mathcal{O},2}^{(1)}[0,l]\big)$ and the 2nd term comes from $\big(a_{\mathcal{O},3}^{(1)}[0,l]+a_{\mathcal{O},4}^{(1)}[0,l]\big)$.\\
	The contribution from the discontinuity for $l=2$ is given by 
	\begin{align}\label{disc ans l=2}
		\left. a_{\cal O}^{(1)}[0,2]\right|_{\rm disc}
		= K_2\frac{  \big(\tilde m+3 \left(e^{\tilde m}-1\right) \log \left(1-e^{-\tilde m}\right)\big)}{96 r^2\left(e^{\tilde m}-1\right) }.
	\end{align}
	For $l=0$,  the result is given by 
	\begin{align}\label{disc ans l=0}
		\left. a_{\cal O}^{(1)}[0,0]\right|_{\rm disc}=	\frac{{K_0} }{12 r^2 \left(e^{\tilde m}-1\right) \tilde m }.
	\end{align}
	
	\subsection*{Arc contribution}
	
	Now we evaluate the  contribution  from the arcs at infinity 
	in the inversion formula.  This  should be added to the contribution from the 
	discontinuity  evaluated in \eqref{disc ans}, \eqref{disc ans l=2} and \eqref{disc ans l=0} to obtain the 1st subleading correction to the one point functions. Thus we need to apply the inversion formula for the arc part given in \eqref{arc} on 1st subleading correction to the correlator $g^{(1)}(z,\bar z)$ given in \eqref{g^1}, as shown below
	\begin{align}\label{arc g 1}
	&	\hat a_{\rm arcs} (\Delta_{},  l)  =  2 K_l   \int_0^1 \frac{dR}{R^{\Delta_{}  +1 - 2\Delta_\phi} }  \times   \nonumber
		 \oint \frac{dw}{i w} \lim_{|w| \rightarrow \infty} 
		\left[ \Big( \frac{ w - w^{-1} }{i} \Big)^{2\nu} 
		F_l(w^{-1}) e^{i\pi\nu}  g^{(1)}( R, w) 
		\right],\\
		&\\
		&{\rm where }\qquad z=R w\qquad {\rm and}\qquad \bar z=R w^{-1}\nonumber.
	\end{align}
	At  $|w|\to \infty$ only $m=0$ term from the correlator \eqref{g^1} survives as rest of terms are exponentially suppressed at large $w$.
	The arc contribution to the inversion formula is non-vanishing for $l\le 4$
	\footnote{This is unlike  the leading large radius term in the 
	 one point function for which it is only $l=0$ contributes from the arc at infinity.}. The reason is that
	the contour integral in $w$ along the circle at infinity has the only non-vanishing contribution from the $w$ independent part of 
	\begin{align}
		\lim_{|w| \rightarrow \infty} 
		\left[ \Big( \frac{ w - w^{-1} }{i} \Big)^{2\nu} 
		F_l(w^{-1}) e^{i\pi\nu}  g^{(1)}( \tilde r, w)\right] .
	\end{align}
	And $w$ independent term exists only for $l\le 4$.
	For $l=4$ we evaluate the arc contribution as the following.
	It is easy to see that $w$ independent part of the above expression is given by
	\begin{align}
		\lim_{|w| \rightarrow \infty} 
		\left[ \Big( \frac{ w - w^{-1} }{i} \Big)^{2\nu} 
		F_l(w^{-1}) e^{i\pi\nu}  g^{(1)}( r, w)\right] =-\frac{1}{96\pi\tilde m r} \frac{\left(\tilde m^2 R^2+\tilde m R\right)}{16} e^{-\tilde m R}\nonumber\\+O(w^{-2}).
	\end{align}
	Thus plugging the $w$ independent term from the above equation into \eqref{g^1} we get
		\begin{align}
		\hat a_{\rm arcs}(\Delta,l)=-\frac{K_l}{24\tilde m r^2}  \int_0^{\infty } \frac{\left(\tilde m^2 R^2+\tilde m R\right) e^{-\tilde m R}}{16 R^{\Delta }} \, dR.
	\end{align}
	Note that contour integral in $w$ along the complete circle at infinity gives $2\pi$ and we have taken this into account in the above equation.
	Now the above integral in $R$ is performed from $0$ to $\infty$ although the originally the upper limit of the integral was till $1$ as it was in \eqref{arc g 1}, because this does not alter the pole structure of $\hat a_{\rm arcs}(\Delta,l)$ in $\Delta$ plane as well the value of the residue at the pole is kept intact. Thus we obtain
	\begin{align}\label{arc l=4}
		a_{\cal O}[0,4]\Big|_{\rm arc}=-\hat a_{\rm arcs}(\Delta,4)|_{\rm Res\ at\ \Delta=5}=-\frac{\tilde m^3}{1152 r^2}K_4.
	\end{align}
	Similarly, for $l=2 $ and $l=0$ respectively we have
	\begin{align}\label{arc l=2}
		a_{\cal O}[0,2]\Big|_{\rm arc}=-\hat a_{\rm arcs}(\Delta,2)|_{\rm Res\ at\ \Delta=3}=\frac{\tilde m}{48 r^2}K_2,
	\end{align}
	and
	\begin{align}\label{arc l=0}
		a_{\cal O}[0,0]\Big|_{\rm arc}=-\hat a_{\rm arcs}(\Delta,0)|_{\rm Res\ at\ \Delta=1}=\frac{1}{24 \tilde m r^2}K_0.
	\end{align}
	
	Now combining the arc part and disc part for $l=0$ given in \eqref{arc l=0} and \eqref{disc ans l=0} respectively gives the 1st subleading correction to the thermal one point function of the operator $\phi^2$.
	Demanding that $\langle\phi^2\rangle_\beta$ vanishes, we recover the gap equation correctly till the first 
	subleading order
	\begin{align}\label{gap }
	{\tilde m}+{2\log \left(1-e^{-\tilde m}\right)}-
		\frac{e^{\tilde m}+1}{24 \left(e^{\tilde m}-1\right) \tilde m r^2}+O\Big(\frac{1}{r^4}\Big)=0.
	\end{align}
	where the leading order term at large $r$ can be read out from the inversion formula as done in 
	\cite{David:2023uya,David:2024naf}. 
	Note that  this agrees with the gap equation obtained from the partition function in \eqref{gapeqorderby}. The gap equation can be solved as an order by order expansion in large $r$ which results in the solution for the 
	thermal mass given by 
	\begin{align}\label{mth large r}
		\tilde m=2 \log\frac{1+\sqrt{5}}{2}+\frac{1}{48 r^2 \csch^{-1} 2}+O\Big(\frac{1}{r^4}\Big).
	\end{align}

For $l=2$ by combining the disc and the arc part given in \eqref{disc ans l=2} and \eqref{arc l=2} respectively we obtain 1st subleading correction to the thermal one point function of the spin-2 current. Along with the one point function of spin-2 current at the leading order evaluated in \cite{David:2023uya,David:2024naf}, we can write down the one point function of spin-2 current till 1st sub-leading correction as
 \begin{align}
 	a_{\cal O}[0,2]=&\frac{-\tilde m^2 \log \left(1-e^{-\tilde m}\right)+3 \tilde m \text{Li}_2\left(e^{-\tilde m}\right)+3 \text{Li}_3\left(e^{-\tilde m}\right)}{6 \pi }\\\nonumber
 	&+\frac{1}{r^2}\Big(\frac{\tilde m}{72 \pi  }+\frac{\tilde m+3 \left(e^{\tilde m}-1\right) \log \left(1-e^{-\tilde m}\right)}{144 \pi  \left(e^{\tilde m}-1\right) }\Big)+O(r^{-4}).
 \end{align}
 
 Now using the large $r$ expansion of the thermal mass $\tilde m$ given in \eqref{mth large r} which satisfies the gap equation \eqref{gap }, in the above equation it is easy to see that the 1st subleading correction at large $r$ to the thermal one point function of the spin-2 current vanishes as given below.
 \begin{align}
 	a_{\cal O}[0,2]=\frac{2 \zeta (3)}{5 \pi }+O\left(r^{-4}\right).
 \end{align}
This is in agreement with the stress tensor calculated from the partition function given in \eqref{energy} where we have seen that there is no non-vanishing correction in the 1st subleading order. Note that we have not obtained 
the perturbative expansion using the inversion formula to the $1/r^4$ order. However the direct expansion of the 
2-point function given in (\ref{larg2ptexp}) can be systematically organised as the expansion in terms of 
Gegenbauer polynomials in (\ref{opexpan}) till order $l =2$ reasonably easily as shown in appendix \ref{A}. From this we have 
verified that the $1/r^4$ term for the stress tensor indeed agrees with (\ref{energy}). 

For $l=4$ combining the disc and arc contributions from \eqref{disc ans} and \eqref{arc l=4} we have the one point function as
 
\begin{align}
a_{\cal O}[0,4]=	-\frac{1}{10080 \pi  (e^{\tilde m}-1) r^2}\Big[\tilde m^2 \left(\left(8 e^{\tilde m}-9\right) \tilde m-18 \left(e^{\tilde m}-1\right) \log \left(1-e^{-\tilde m}\right)\right)\nonumber\\
+81 \left(e^{\tilde m}-1\right) \tilde m \text{Li}_2(e^{-\tilde m})+81 \left(e^{\tilde m}-1\right) \text{Li}_3(e^{-\tilde m})\Big]
\end{align}
\\
 The 1st subleading correction to the thermal one point function of ${\cal O}[0,l]$ at the Gaussian fixed point can be obtained by taking $\tilde m\to 0$ limit in the equation \eqref{disc ans} which gives the following answer
 \begin{align}
 	a_{\cal O}^{\rm free{(1)}}[0,l]=-K_l\frac{\left(l^4-2 l^3-7 l^2+20 l-15\right) \Gamma \left(l-\frac{5}{2}\right)}{192 \sqrt{\pi } r^2 \Gamma (l+1)} \zeta (l-1),\quad{\rm for }\quad {\rm even}\ l\ge 4.
 \end{align}
 We restrict $\tilde m\to 0$ limit to operators of spin $l\ge 4$ since for $l=0,2$ this limit is singular indeed the partition function after the $\frac{\beta}{r}$ expansion in  \eqref{log Z} is singular. 

	\subsection*{Large $l$ limit}
\begin{figure} 
\begin{center} 
		\includegraphics[scale=.58] {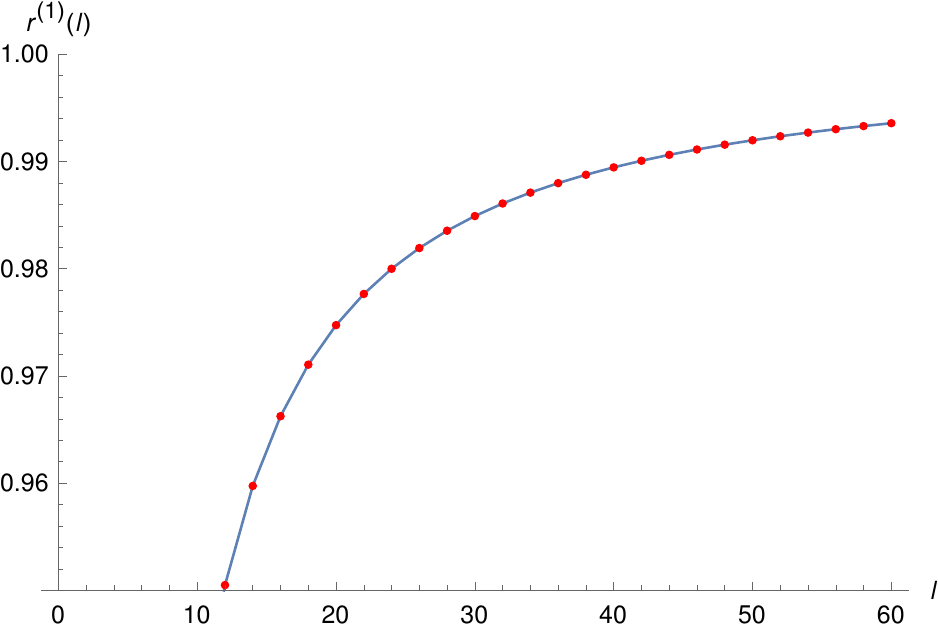}
		\caption{The ratio of the  leading finite size correction  of the spin-$l$ one point function 
		of the critical theory to that of the 
		free theory defined in (\ref{ratio}) versus the spin $l$. } \label{large l}
	\end{center}
\end{figure}

	In \cite{David:2023uya,David:2024naf}, we have shown that thermal one point functions of operators ${\cal O}[0,l] $ on $S^1_\beta\times R^{d-1}$ evaluated at the non-trivial fixed point of the theory tend to the thermal one point functions evaluated at the free CFT point at large spin-$l$ limit of the operators. Now  that we have evaluated  
	 the thermal one point functions on $S^1_\beta\times S^2$ to the leading $1/r^2$ order it is interesting to 
	 check if these corrections also exhibit the same behaviour in the large spin limit. 
	To study this we consider the ratio of  the thermal one point function computed at the critical point or the non-trivial fixed point of the theory to that at the free CFT point as defined below
	\begin{align}
		r(l)=\frac{a_{\cal O}[0,l]}{a_{\cal O}^{\rm free}[0,l]},
	\end{align}
%
	We evaluate the thermal one point function of the operator ${\cal O}[n,l]$ till the 1st subleading correction at large $r$ by combining the thermal one point function at  leading order obtained in \cite{David:2024naf}\footnote{We use the result obtained in equation (2.53) of \cite{David:2024naf} at zero chemical potential $\mu=0$. $a_{\cal O}^{\rm free {(0)}}[0,l]$ is obtained by taking both $\mu=m_{\rm th}=0 $ in this equation.} with the 1st subleading correction obtained in \eqref{disc ans}, then substituting the thermal mass \eqref{mth large r} in it, we finally organise in orders of $\frac{1}{r}$ as,
	\begin{align}
		a_{\cal O}[0,l]=\tilde a_{\cal O}^{(0)}[0,l]+\frac{1}{r^2}\tilde a_{\cal O}^{(1)}[0,l]+O(r^{-4}).
	\end{align}
	where $ \tilde a_{\cal O}^{(0)}[0,l]$ and $\tilde a_{\cal O}^{(1)}[0,l]$ are constants depending on $l$ 
	determined with the use of the gap equation.
	For the free theory as well we have the expansion
	\begin{align}
		a_{\cal O}^{\rm free}[0,l]= a_{\cal O}^{(0){\rm free}}[0,l]+\frac{1}{r^2} a_{\cal O}^{(1){\rm free}}[0,l]+O(r^{-4}),
 	\end{align}
	where 
	\begin{align}
		a_{\cal O}^{(0){\rm free}}[0,l]=\frac{\zeta(l+1)}{2 \pi },\qquad {\rm for}\ l\ {\rm even.}
	\end{align}
	\begin{align}
		a_{\cal O}^{\rm free{(1)}}[0,l]=-K_l\frac{\left(l^4-2 l^3-7 l^2+20 l-15\right) \Gamma \left(l-\frac{5}{2}\right)}{192 \sqrt{\pi } r^2 \Gamma (l+1)} \zeta (l-1),\qquad{\rm for }\qquad {\rm even}\ l\ge 4.
	\end{align}
	In \cite{David:2023uya,David:2024naf} it has been shown that
		\begin{align}
		\lim_{l\to \infty}r^{(0)}(l)=\frac{\tilde a_{\cal O}^{(0)}[0,l]}{a_{\cal O}^{\rm free(0)}[0,l]}\to 1.
	\end{align}
	We now numerically evaluate the  ratio of the 1st subleading correction to the thermal one point function at the non-trivial fixed point to that at the free CFT defined by 
	\begin{align}
		\lim_{l\to \infty}r^{(1)}(l)=\frac{\tilde a_{\cal O}^{(1)}[0,l]}{a_{\cal O}^{\rm free(1)}[0,l]}.
	\end{align}
	The result is shown in  figure \ref{large l}. We observe that 
	\begin{align} \label{ratio}
		\lim_{l\to \infty}r^{(1)}(l)=\frac{\tilde a_{\cal O}^{(1)}[0,l]}{a_{\cal O}^{\rm free(1)}[0,l]}\to 1.
	\end{align}
	This implies that the ratio $r(l)$ tends to $1$ on including the first finite size correction. 
	This  observation provides more evidence that thermal observables at large spin tend to 
	their free values. 

\section{Conclusions} \label{sec6}

We have evaluated  the leading corrections to the 
 thermal one point functions of higher spin bi-linears of the  critical $O(N)$ model on $S^1\times S^2$ 
in the $\beta/r$ expansion in the large $N$ limit.
For this we obtained the partition function of a massive scalar  in (\ref{log Z}) 
 as well as its 2 point function in position space 
on $S^1\times S^2$ in (\ref{massivecorr1}) .  
As far as we are aware the expressions for these quantities  derived in this  paper 
 are new.  They are of the 
form that allow a power series expansion in terms of $\beta/r$.  To obtain these expressions 
we used methods developed to express partition functions of free theories as integrals over 
Harish-Chandra characters. 

We developed a perturbative expansion of the Euclidean inversion formula and obtained the leading finite 
size corrections to the one point functions of higher spin currents of the $O(N)$ model on $S^1\times S^2$ at large $N$. 
The finite size corrections to the one point functions 
also demonstrated the property seen in \cite{David:2023uya,David:2024naf} for the leading order. 
These corrections tend to that of the free theory at the Gaussian fixed point in the large spin limit.

The methods developed in this paper can be generalized to fermions and  to other dimensions. 
A systematic  study of these corrections can be made so as to find what information they provide for the
expansion of the one point functions in terms of conformal blocks on $S^1\times S^2$.  It is also possible to extend the 
methods developed here to other geometries namely $S^1 \times AdS_2$ and the squashed sphere studied in 
\cite{Hartnoll:2005yc}. 

Since we have a method to obtain the partition function of the critical  $O(N)$ at large $N$ as a power series 
in $\beta/r$, it will be interesting to study it in detail to see if there are signatures of phase transitions 
that can occur at some critical value of $\beta/r$.  If the model does exhibit a phase transition, then 
it would be remarkable and the implications of this observation to the holographic dual can be explored.

\acknowledgments

We thank Subir Sachdev for discussions and correspondence after the first version of the paper appeared on the 
arXiv which led us to evaluate the finite size corrections to the density of states. 
S.K would like to thank  Volker Schomerus and the ``Theory'' group at DESY, Hamburg for discussions, 
and an opportunity to present research leading up to this work. 
S.K would like to thank Raghu Mahajan,  Ashok Sen and the String theory group at ICTS-TIFR Bengaluru for their valuable comments during the presentation of the work precursor to this. 
S.K would like to thank the organisers of the autumn school on "Quantum Integrable Models" the CRC 1624 "Higher structures, moduli spaces, and integrability" in Hamburg for the warm hospitality during the final stage of this work.

\appendix
\section{OPE from the direct expansion of the 2-point function}
\label{A}
We expand the 2-point function \eqref{larg2ptexp} as a taylor series in small $\sqrt{\tau^2+\tilde \theta^2}$ up to a few orders and identify first few coefficients of the Gegenbauer polynomials as organised in the equation \eqref{opexpan}. Here we find the next subleading correction to the gap equation  and the stress tensor using this OPE expansion and show that it agrees with those evaluated from the partition function. This agreement supports the fact that $\sqrt{\tau^2+\tilde \theta^2}$ plays the role of the length correctly in the OPE expansion in the $S^1\times S^2$ geometry which has been used in our study of the inversion formula.
 	{ \begin{align} 
		&\hat g(\tau,\tilde \theta)\nonumber\\
		&=-\frac{\tilde m }{4\pi } 
		\Bigg[-\frac{e^{-\tilde m \sqrt{\tilde \theta ^2+\tau ^2}}}{ \tilde m \sqrt{\tilde \theta ^2+\tau ^2}}+\frac{e^{- \tilde m \sqrt{\tilde \theta ^2+\tau ^2}} \left(-(\tilde \theta ^2+\tau ^2)^{3/2}+\tilde \theta ^4  \tilde m^2 \sqrt{\tilde \theta ^2+\tau ^2}-\tilde m (2 \tilde \theta ^2 \tau ^2+\tilde \theta ^4)\right)}{24 \tilde m^2 r^2 (\tilde \theta ^2+\tau ^2)^{3/2}}\nonumber\\
		&+\frac{1}{r^4} \frac{e^{- \tilde m \sqrt{\tilde \theta ^2+\tau ^2}} }{5760  \tilde m^4 (\tilde \theta ^2+\tau ^2)^{5/2}}\Big(-21 (\tilde \theta ^2+\tau ^2)^{5/2}-36 \tilde \theta ^2  \tilde m^2 (\tilde \theta ^2+\tau ^2)^{5/2}-5\tilde  \theta ^8  \tilde m^5 (\tilde \theta ^2+\tau ^2)\nonumber\\
		&- \tilde m^3 (80 \tilde \theta ^6 \tau ^2+66 \theta ^4 \tau ^4+29\tilde  \theta ^8)+\tilde \theta ^6  \tilde m^4 (37 \tilde \theta ^2+52 \tau ^2) \sqrt{\tilde \theta ^2+\tau ^2}-21  \tilde m (\tilde \theta ^2+\tau ^2)^3\Big)+O\Big(\frac{1}{r^6}\Big)\Bigg],\nonumber\\
		\equiv& \hat g^{(0)}(\tau,\tilde\theta)+\hat g^{(1)}(\tau,\tilde \theta)+\hat g^{(2)}(\tau,\tilde\theta) +O\Big(\frac{1}{r^6}\Big).
\end{align}}
Recall that $ \hat g^{(0)}(\tau,\tilde \theta)$ is the flat space correlator and $\hat g^{(1)}(\tau,\tilde \theta), \hat g^{(2)}(\tau,\tilde \theta)$ are the leading and 1st sub-leading finite size  corrections to the correlator respectively. First we expand $\hat g^{(0)}(\tau+m, \tilde \theta)$ and $\hat g^{(0)}(\tau,\tilde \theta) $ separately as we need to perform the sum over images as it was shown  in equation \eqref{imagsum}.
\begin{align}\label{g0 m}
\hat g^{(0)}(\tau+m,\theta)=-\frac{\tilde m}{4\pi}\Bigg[&-\frac{e^{-\tilde m | m| }}{\tilde m | m| }+\frac{\eta  |x| e^{-\tilde m | m| } (\tilde m | m| +1)}{m \tilde m | m| } \nonumber\\
&+	\frac{|x|^2 e^{-\tilde m | m| } \left(\eta ^2 \left(-\tilde m^2\right) | m| ^2+| m|  \left(\tilde m-3 \eta ^2 \tilde m\right)-3 \eta ^2+1\right)}{2 m^2 \tilde m | m| }+\cdots\Bigg].
\end{align}
Recall that $|x|=\sqrt{\tau^2+\tilde \theta^2}$ and $\eta=\frac{\tau}{|x|}$.
\begin{align}\label{g0 0}
\hat g^{(0)}(\tau,\theta)=-\frac{\tilde m}{4\pi}\Bigg[-\frac{1}{\tilde m |x|}+1	-\frac{\tilde m |x|}{2}+\frac{\tilde m^2 |x|^2}{6}+\cdots\Bigg].
\end{align}
The expansion for $\hat g^{(1)}(\tau+m,\theta)$ and $\hat g^{(1)}(\tau,\theta)$ is given below
\begin{align}\label{g1 m}
\hat g^{(1)}(\tau+m,\theta)=&-\frac{\tilde m}{4\pi}\Bigg[	-\frac{e^{-\tilde m | m| }}{24 \left(\tilde m^2 r^2\right)}+\frac{\eta  x | m|  e^{-\tilde m | m| }}{24 m \tilde m r^2}\nonumber\\
&+\frac{x^2 e^{-\tilde m | m| } \left(-\left(\eta ^2-1\right) | m| ^4+\eta ^2 (-\tilde m) | m| ^5+4 \left(\eta ^2-1\right) m^4\right)}{48 m^2 \tilde m r^2 | m| ^3}+\cdots\Bigg].
\end{align}
For $ m=0$
\begin{align}\label{g1 0}
	\hat g^{(1)}(\tau,\tilde\theta)=-\frac{\tilde m}{4\pi}\Bigg[-\frac{1}{24 \left(\tilde m^2 r^2\right)}+\frac{\eta ^4 x}{24 \tilde m r^2}+\frac{\left(3-4 \eta ^2\right) x^2}{48 r^2}+\cdots\Bigg].
\end{align}
The expansion for $\hat g^{(2)}(\tau+m,\theta)$ and $\hat g^{(2)}(\tau,\theta)$ is given below
\begin{align}\label{g2 m}
\hat g^{(2)}(\tau+m,\tilde\theta)=	-\frac{\tilde m}{4\pi}\Bigg[-\frac{7 \left(e^{-\tilde m | m| } (\tilde m | m| +1)\right)}{1920 \tilde m^4}+\frac{7 \eta  |x| | m| ^2 e^{-\tilde m | m| }}{1920 m \tilde m^2}\nonumber\\
+\frac{|x|^2 e^{-\tilde m | m| } \left(-7 \eta ^2 \tilde m | m| ^3+7 | m| ^2+24 \left(\eta ^2-1\right) m^2\right)}{3840 m^2 \tilde m^2}+\cdots\Bigg].
\end{align}
\begin{align}\label{g2 0}
\hat g^{(2)}(\tau,\tilde\theta)=-\frac{\tilde m}{4\pi}\Bigg[	\frac{\left(24 \eta ^2-17\right) |x|^2}{3840 \tilde m^2}-\frac{7}{1920 \tilde m^4}+\cdots\Bigg].
\end{align}
Finally we combine all these expansions \eqref{g0 m},\eqref{g0 0},\eqref{g1 m},\eqref{g1 0},\eqref{g2 m} and \eqref{g2 0} as shown in the following
\begin{align}
	g(\tau, \theta) = \sum_{m=-\infty}^\infty \hat g ( \tau + m, \theta) 
	=\sum_{m=-\infty}^\infty  [\hat g^{(0)}(\tau+m,\tilde\theta)+\hat g^{(1)}(\tau+m,\tilde \theta)&+\hat g^{(2)}(\tau+m,\tilde\theta)] \nonumber\\&+O\Big(\frac{1}{r^6}\Big).
\end{align}
Performing the sum over images we organise it as the linear sum of Gegenbauer polynomials to  obtain 
\begin{align}
	g(\tau,\tilde \theta)=-\frac{\tilde m}{4\pi } \Bigg[\Big(-\frac{e^{\tilde m}+1}{24 \left(e^{\tilde m}-1\right) \tilde m^2 r^2}-\frac{7 \left(2 e^{\tilde m} \tilde m+e^{2 \tilde m}-1\right)}{1920 \left(e^{\tilde m}-1\right)^2 \tilde m^4 r^4}+\frac{2 \log \left(1-e^{-\tilde m}\right)}{\tilde m}+1\Big)\nonumber\\
	+\frac{2}{3}|x|^2 C_2^{(\frac{1}{2})}(\eta) \Big(\frac{\tilde m^2 \log \left(e^{-\tilde m} \left(e^{\tilde m}-1\right)\right)-3 \tilde m \text{Li}_2\left(e^{-\tilde m}\right)-3 \text{Li}_3\left(e^{-\tilde m}\right)}{\tilde m}\nonumber\\
	-\frac{\left(2 e^{\tilde m}-1\right) \tilde m+3 \left(e^{\tilde m}-1\right) \log \left(1-e^{-\tilde m}\right)}{24 \left(e^{\tilde m}-1\right) \tilde m r^2}
	+\frac{-14 e^{\tilde m} \tilde m+24 e^{2 \tilde m}-24}{(3840 \left(e^{\tilde m}-1\right)^2 \tilde m^2) r^4}\Big)+\cdots\Bigg].
\end{align}
From the coefficient of $|x|^0 C_0^{(\frac{1}{2})}(\eta)$ we recover the gap equation correctly till the order $\frac{1}{r^4}$,
 \begin{align}
\tilde m+{2 \log \left(1-e^{-\tilde m}\right)}	-\frac{e^{\tilde m}+1}{24 \left(e^{\tilde m}-1\right) \tilde m r^2}-\frac{7 \left(2 e^{\tilde m} \tilde m+e^{2 \tilde m}-1\right)}{1920 \left(e^{\tilde m}-1\right)^2 \tilde m^3 r^4}+O(\frac{1}{r^6})=0.
\end{align}
This agrees with the gap equation \eqref{gapeqorderby} obtained from the partition function and as shown earlier in \eqref{gap soln} the gap equation has the solution given below.
\begin{align}\label{gap soln 1}
	\tilde m=2 \log\frac{1+\sqrt{5}}{2}+\frac{1}{r^2}\frac{1}{48 \text{csch}^{-1}2} +\frac{1}{r^4}\frac{55+64 \sqrt{5} \text{csch}^{-1}2}{230400 (\text{csch}^{-1} 2)^3}+O\big(\frac{1}{r^6}\big).
\end{align}
The coefficient of $|x|^2 C_2^{(\frac{1}{2})}(\eta)$ gives the spin-2 current $a_{\cal O}[0,2]$ to be 
\begin{align}
	a_{\cal O}[0,2]= & \frac{2}{3} \Big(\frac{\tilde m^2 \log \left(e^{-\tilde m} \left(e^{\tilde m}-1\right)\right)-3 \tilde m \text{Li}_2\left(e^{-\tilde m}\right)-3 \text{Li}_3\left(e^{-\tilde m}\right)}{\tilde m}\nonumber\\
	&-\frac{\left(2 e^{\tilde m}-1\right) \tilde m+3 \left(e^{\tilde m}-1\right) \log \left(1-e^{-\tilde m}\right)}{24 \left(e^{\tilde m}-1\right) \tilde m r^2}
	+\frac{-14 e^{\tilde m} \tilde m+24 e^{2 \tilde m}-24}{(3840 \left(e^{\tilde m}-1\right)^2 \tilde m^2) r^4}\Big)+O(\frac{1}{r^6}).
\end{align}
Now substituting the thermal mass \eqref{gap soln 1}  in the above equation we obtain the spin-2 current at the critical point of the theory on $S^1\times S^2$ to be 
\begin{align}
	a_{\cal O}[0,2]=\frac{2 \zeta (3)}{5 \pi }-\frac{1}{576 \sqrt{5} \pi  r^4 \text{csch}^{-1}(2)}+O(\frac{1}{r^6}).
\end{align}
And this agrees with the energy density evaluated from the partition function obtained at \eqref{energy} up to an overall constant.
\section{An alternate approach to apply the inversion formula}
\label{B}
In this appendix we apply the inversion formula \eqref{a disc} directly on the 1st line of the equation \eqref{int 1st term} without rewriting it as an integral over $\tilde m$ as was done in the 2nd line of \eqref{int 1st term} and show that it reproduces the same result obtained using the latter.\\ Applying the inversion formula \eqref{a disc} on the 1st line of \eqref{int 1st term} we have
\begin{align}
\hat a_{{\cal O},1}[n,l]|_{\rm Disc}=	\frac{K_l \tilde m}{96 \pi r^2}\int_0^1 \frac{d\bar z}{\bar z}\int_m^{{\rm max}(m,\frac{1}{\bar z})}\frac{d z}{ z}\frac{(z-\bar z)^5 F_l\left(\sqrt{\frac{\bar z}{z}}\right)  \sin \left(m_{\rm th} \sqrt{(z-m) (m-\bar z)}\right) }{8 (z\bar z)^{\Delta/2} (z-m) (m-\bar z)}.
\end{align}
We use the following coordinate transformation
\begin{align}
	\bar z=z z',\qquad \qquad{\rm and}\qquad z=m z'.
\end{align}
to obtain 
\begin{align}
&\hat a_{{\cal O},1}[n,l]|_{\rm Disc}
=	\frac{K_l \tilde m}{96 \pi r^2}\times\nonumber\\
&\int_0^1 d\bar z\int_1^{{\rm max}(1,\frac{1}{m\sqrt{\bar z}})} dz\frac{m^3 z^4 (\bar z-1)^5 \bar z^{\frac{l-1}{2}} \, _2F_1\left(\frac{1}{2},l+1;l+\frac{3}{2};\bar z\right)  \sin \left(m \tilde m \sqrt{(1-z) (z \bar z-1)}\right)}{8\left(m^2 z^2 \bar z\right)^{\frac{\Delta }{2}} (z-1) (z \bar z-1)}.
\end{align}
Now expanding the above expression in small $\bar z$ and keeping only the leading term and performing the $\bar z$ integral we have
\begin{align}
\hat a_{{\cal O},1}^{(1)}[0,l]|_{\rm Disc}=	\frac{K_l \tilde m}{96 \pi r^2}\int_1^\infty dz\frac{m^3 z^{\frac{l+9}{2}} \left(m^2 z^3\right)^{-\frac{\Delta }{2}} \sin \left(m \tilde m \sqrt{z-1}\right)}{4 (z-1) (-\Delta +l+1)}.
\end{align}
Changing the integration variable $z=1+z'^2$,
\begin{align}
\hat a_{{\cal O},1}^{(1)}[0,l]|_{\rm Disc}=	\frac{K_l \tilde m}{96 \pi r^2}\int_1^\infty dz \frac{m^{3-\Delta } \left(z^2+1\right)^{\frac{1}{2} (-3 \Delta +l+9)} \sin (m \tilde m z)}{2 z (-\Delta +l+1)}.
\end{align}
The negative  residue at $\Delta=l+1$ of the above expression is given by
 \begin{align}\label{direct ans}
 a_{{\cal O},1}^{(1)}[0,l]|_{\rm Disc}=-\frac{K_l \tilde m}{96 \pi r^2}	\sum_{m=1}^\infty\frac{m^{2-l}}{4 \Gamma (l-3)} \Bigg(-\Gamma \left(l-\frac{7}{2}\right) \sqrt{\pi } (m \tilde m) \, _1F_2\left(\frac{1}{2};\frac{3}{2},\frac{9}{2}-l;\frac{m^2 \tilde m^2}{4}\right)\nonumber\\+\pi  \Gamma (l-3)
	+ \frac{2\pi \Gamma (6-2 l)}{\Gamma(4-l)}  (m \tilde m)^{2l-6}  \, _1F_2\left(l-3;l-\frac{5}{2},l-2;\frac{m^2 \tilde m^2}{4}\right)
	\Bigg).\nonumber\\
\end{align}
Now this expression agrees with \eqref{a_1} for various values of $l$ as shown below. From \eqref{direct ans} we have 
\begin{align}
	a_{{\cal O},1}^{(1)}[0,0]=&a_{{\cal O},1}^{(1)}[0,2]=0,\\
	a_{{\cal O},1}^{(1)}[0,4]=&-K_4\frac{\tilde m \text{Li}_2\left(e^{-\tilde m}\right)}{384 r^2},\\
a_{{\cal O},1}^{(1)}[0,6]=&-K_6\frac{5 \tilde m^2 \text{Li}_3\left(e^{-\tilde m}\right)+\tilde m^3 \text{Li}_2\left(e^{-\tilde m}\right)+8 \tilde m \text{Li}_4\left(e^{-\tilde m}\right)}{3072 r^2}\nonumber,\\
a_{{\cal O},1}^{(1)}[0,8]=&-\frac{K_8}{147456 r^2}\nonumber
\big(14 \tilde m^4 \text{Li}_3(e^{-\tilde m})+87 \tilde m^3 \text{Li}_4(e^{-\tilde m})+279 \tilde m^2 \text{Li}_5(e^{-\tilde m})\\&\qquad\qquad\qquad\qquad\qquad\qquad+\tilde m^5 \text{Li}_2(e^{-\tilde m})+384 \tilde m \text{Li}_6(e^{-\tilde m})\big)\nonumber,\\
a_{{\cal O},1}^{(1)}[0,10]=&- \frac{K_{10}}{17694720 r^2} \big(27 \tilde m^6 \text{Li}_3(e^{-\tilde m})+345 \tilde m^5 \text{Li}_4(e^{-\tilde m})+2640 \tilde m^4 \text{Li}_5(e^{-\tilde m})\nonumber\\
&+12645 \tilde m^3 \text{Li}_6(e^{-\tilde m})+35685 \tilde m^2 \text{Li}_7(e^{-\tilde m})+\tilde m^7 \text{Li}_2(e^{-\tilde m})+46080 \tilde m \text{Li}_8(e^{-\tilde m})\big)\nonumber.
\end{align}
These exactly match with the formula \eqref{a_1}.\\

Similarly we use the inversion formula \eqref{a disc} on $g_2(z,\bar z)$ from \eqref{g^1} and show that it agrees with  \eqref{a_2} obtained by rewritting $g_2(z,\bar z)$ as integral over $\tilde m$ twice and using the the inversion formula \eqref{a disc} on the integrand and finally performing the two $\tilde m$ integrals. 
The result obtained by using the inversion formula \eqref{a disc} directly on $g_2(z,\bar z)$ from \eqref{g^1} by using the similar method as used above is given by 
\begin{align}
&a_{{\cal O},2}^{(1)}[0,l]=	\frac{1}{
		384\pi r^2  \tilde m^4 }   \sum_{m=1}^\infty
		m^{-l-4} \Bigg[-2 \sqrt{\pi } m^5 \tilde m^5  \frac{\Gamma\left(l-\frac{5}{2}\right)}{\Gamma
		(l-3)} \, _1F_2\left(-\frac{1}{2};\frac{1}{2},\frac{7}{2}-l;\frac{m^2 \tilde m^2}{4}\right)\nonumber\\
		&-2 \sin (\pi  (4-l)) \Gamma (5-2 l)  (m \tilde m)^{2 l} \, _1F_2\left(l-3;l-2,l-\frac{3}{2};\frac{m^2 \tilde m^2}{4}\right)+\pi  m^6 \tilde m^6 \Bigg].
\end{align}
For various integer values of $l$ we have
\begin{align}
	a_{{\cal O},2}^{(1)}[0,0]&=a_{{\cal O},2}^{(1)}[0,2]=0,\\
	a_{{\cal O},2}^{(1)}[0,4]&=	-K_4\frac{\tilde m \text{Li}_3\left(e^{-\tilde m}\right)}{384 r^2},\nonumber\\
	a_{{\cal O},2}^{(1)}[0,6]&=-K_6\frac{\tilde m \left(\tilde m^2 \text{Li}_3(e^{-\tilde m})+7 \tilde m \text{Li}_4\left(e^{-\tilde m}\right)+15 \text{Li}_5(e^{-\tilde m})\right)}{3072 r^2}\nonumber,\\
	a_{{\cal O},2}^{(1)}[0,8]&=-K_8\frac{\tilde m }{147456 r^2} \big(\tilde m^4 \text{Li}_3(e^{-\tilde m})+18 \tilde m^3 \text{Li}_4(e^{-\tilde m})+141 \tilde m^2 \text{Li}_5(e^{-\tilde m})\nonumber\\
	&
	\qquad\qquad\qquad\qquad\qquad
	+561 \tilde m \text{Li}_6(e^{-\tilde m})+945 \text{Li}_7(e^{-\tilde m})\big)\nonumber,\\
	a_{{\cal O},2}^{(1)}[0,10]&=-K_{10}\frac{\tilde m }{17694720 r^2} \big(\tilde m^6 \text{Li}_3(e^{-\tilde m})+33 \tilde m^5 \text{Li}_4(e^{-\tilde m})+510 \tilde m^4 \text{Li}_5(e^{-\tilde m})\nonumber\\
	&+4680 \tilde m^3 \text{Li}_6(e^{-\tilde m})+26685 \tilde m^2 \text{Li}_7(e^{-\tilde m})+89055 \tilde m \text{Li}_8(e^{-\tilde m})+135135 \text{Li}_9(e^{-\tilde m})\big)\nonumber.
\end{align}
These agree with the  formula \eqref{a_2} for $a_{{\cal O},2}^{(1)}[0,l]$ for any general $l$.
\section{The partition function at $\tilde m \to 0$ limit}
\label{C}
In this appendix we take  the massless limit of the partition function evaluated for the massive scalar on a 2-sphere obtained in \eqref{part fn order by order}. We show that the partition function at the massless limit agrees with the existing result for the partition function  evaluated directly for a massless scalar on a 2-sphere \cite{Benjamin:2023qsc}. From \eqref{part fn order by order} we have the partition function for the massive scalar to be
\begin{align}
	\log Z\Big(\tilde m,\frac{\beta}{r}\Big)=\frac{r^2}{\beta^2}\Big[\frac{1}{3} (\tilde m\beta)^3+{2 \tilde m \beta }\text{Li}_2(e^{-\tilde m \beta })+{2 }\text{Li}_3(e^{-\tilde m \beta })\Big]-\Big(\frac{\beta \tilde m}{24}+\frac{1}{12} \log (1-e^{-\tilde m\beta})\Big)\nonumber\\
	+\frac{\beta^2}{r^2} \Big(\frac{7 \left(e^{\beta  \tilde m}+1\right)}{1920 \beta  \tilde m \left(e^{\beta  \tilde m}-1\right)}\Big)+\frac{\beta^4}{r^4}\Big(\frac{31 \left(2 \beta  \tilde m e^{\beta  \tilde m}+e^{2 \beta  \tilde m}-1\right)}{64512 \beta ^3 \tilde m^3 \left(e^{\beta  \tilde m}-1\right)^2}\Big)+O\Big(\frac{\beta^6}{r^6}\Big).
\end{align}
Now taking $\tilde m\to 0$ limit,
\begin{align}\label{lim m 0}
	\lim_{\tilde m \to 0}	\log Z\Big(\tilde m,\frac{\beta}{r}\Big)=&\frac{2 r^2 \zeta (3)}{\beta ^2}+
	\Big(\frac{7 \beta ^2}{11520 r^2}+\frac{31 \beta ^4}{11612160 r^4}+\cdots\Big)\nonumber\\
	&+\frac{1}{2}\Big(-\frac{1}{6} \log (\beta  \tilde m)+\frac{7}{480 \tilde m^2 r^2}+\frac{31}{8064 \tilde m^4 r^4}+\cdots\Big).
\end{align}
The 1st term is the flat space answer for the partition function,  the terms inside the first parenthesis are the corrections to the flat space answer in orders of $\frac{\beta}{r}$ and  inside the last parenthesis there is an infinite series of divergent contributions at $\tilde m\to 0$ limit. But we will show that such infinite number of divergent terms combine to end up in a finite contribution and thus our $\tilde m\to 0$ limit is justified.
The terms inside the last parenthesis except $-\frac{1}{6}\log \beta\tilde m$ of the above equation can be written as the following using the general expression given in \eqref{allorderp}
\begin{align}\label{original}
I=	\sum_{p=2}^\infty\frac{(-1)^{p+1} 4^{-p} \left(4^p-2\right)  x^{2p-2}B_{2 p}}{(p-1) p},
\end{align}
where $B_{2p}$ is the Bernoulli number and $x=\frac{1}{mr}$.\\
To evaluate this series we use the method of Borel sum as follows
\begin{align}\label{laplace tr}
	I=\int_0^\infty e^{-t} t^2 f(x t),
\end{align}
where $f(x)$ is the Borel transform of the original series \eqref{original} as given below
\begin{align}\label{f(x)}
	f(x)=\sum_{p=2}^\infty \frac{(-1)^{p+1} (2^{2p-1}-1)2^{-2p+1}x^{2p-2}B_{2p}}{p(p-1)(2p)!}.
\end{align}
Given that $f(x) $ is defined as above, the following quantity has the infinite sum representation which can be recognised as the standard Lorentz series of $\csc \frac{x}{2}$ keeping the 1st two leading terms aside at large $x$ from this Lorentz series as shown below
\begin{align}
	\partial_x(x^3\partial_xf(x))&=2\Bigg(\sum_{p=0}^\infty \frac{(-1)^{p+1}}{(2p)!}B_{2p } 2(2^{2p-1}-1)\Big(\frac{x}{2}\Big)^{2p-1}-\frac{2}{x}-\frac{x}{2}\Bigg)\\
	&=2\Big(\csc \Big(\frac{x}{2}\Big)-\frac{2}{x}-\frac{x}{12}\Big).
\end{align}
Integrating the above expression once w.r.t $x$, we have
\begin{align}
	x^3 f'(x)=-\frac{x^2}{12}-4 \log x+4\log\Big( \tan \frac{x}{4}\Big)+c,
\end{align}
where $c$ is the integration constant. Another integration gives $f(x)$ to be
\begin{align}\label{after int}
	f(x)=\frac{1}{x^2}+\frac{2 \log (x)}{x^2}-\frac{\log (x)}{12}+\frac{c}{x^2}+4 \int \frac{dx}{x^3} \log\Big(\tan \frac{x}{4}\Big)+c',
\end{align}
$c'$ is constant of integration.
Now  using integration by parts twice we have,
\begin{align}\label{tan x/4}
	4\int \frac{dx}{x^3}\log \Big(\tan \frac{x}{4}\Big)=-\frac{2}{x^2} \log \Big(\tan\frac{x}{4}\Big)-\frac{1}{x\csc\frac{x}{2}}+2\int\frac{dx}{x} \frac{d^2}{dx^2} \log \Big(\tan \frac{x}{4}\Big).
\end{align}
And assuming a small imaginary part in $x$ we can write 
\begin{align}
	\log \Big(\tan \frac{x}{4}\Big)=-\sum_{n=0}^\infty \frac{\cos[(n+\frac{1}{2})x]}{n+\frac{1}{2}}.
\end{align}
Substituting this expression in \eqref{tan x/4} we get
\begin{align}
	4\int \frac{dx}{x^3}\log \Big(\tan \frac{x}{4}\Big)&=-\frac{2}{x^2} \log \Big(\tan\frac{x}{4}\Big)-\frac{1}{x\csc\frac{x}{2}}+\int \frac{dx}{x}\sum_{n=0}^\infty (2n+1) \cos [(n+\frac{1}{2} ) x]\nonumber\\
	&=-\frac{2}{x^2} \log \Big(\tan\frac{x}{4}\Big)-\frac{1}{x\csc\frac{x}{2}}+\sum_{n=0}^\infty (2n+1) \text{Ci}\Big((n+\frac{1}{2}) x\Big).
\end{align}
Thus using the above equation in \eqref{after int}
\begin{align}\label{f(x) 2}
	f(x)=\frac{1}{x^2}+\frac{2 \log (x)}{x^2}-\frac{\log (x)}{12}+\frac{c}{x^2}-\frac{2}{x^2} \log \Big(\tan\frac{x}{4}\Big)-\frac{1}{x\sin\frac{x}{2}}\nonumber\\
	+\sum_{n=0}^\infty (2n+1) \text{Ci}\Big((n+\frac{1}{2}) x\Big)+c'.
\end{align}
We have from the original expression of $f(x)$ as was given in \eqref{f(x)}
\begin{align}
	f(0)=0
\end{align}
Using this condition we can fix both $c$ and $c'$ to be
\begin{align}
	c=1-4\log 2\qquad c'=\frac{1}{8}-\frac{\gamma}{12}-\frac{1}{12}\log 2-\zeta'(-1).
\end{align}
Putting the value of  $c$ and $c'$ from the above equation into \eqref{f(x) 2} we get
\begin{align}
	f(x)=\frac{2-4\log 2}{x^2}+\frac{2 \log (x)}{x^2}-\frac{\log (x)}{12}+\frac{1}{8}-\frac{\gamma}{12}-\frac{1}{12}\log 2-\zeta'(-1)\nonumber\\
	+\sum_{n=0}^\infty (2n+1) \text{Ci}\Big((n+\frac{1}{2}) x\Big)	-\frac{2}{x^2} \log \Big(\tan\frac{x}{4}\Big)-\frac{1}{x\sin\frac{x}{2}}.
\end{align}
Finally we should perform the following Laplace transform as was given in \eqref{laplace tr}
\begin{align}
I(x)=	\int_0^\infty f(x t) e^{-t} t^2\equiv I_1(x)+I_2(x)+I_3(x),
\end{align}
where 
\begin{align}
	I_1(x)&=\int_0^{\infty } e^{-t} t^2 dt\Big(\frac{2-4\log 2}{x^2t^2}+\frac{2 \log (xt)}{x^2t^2}-\frac{\log (xt)}{12}+\frac{1}{8}-\frac{\gamma}{12}-\frac{1}{12}\log 2-\zeta'(-1)\Big)\nonumber\\
	I_2(x)&=\int_0^{\infty } e^{-t} t^2 dt \sum_{n=0}^\infty (2n+1) \text{Ci}\Big((n+\frac{1}{2}) xt\Big)\nonumber\\
	I_3(x)&=\int_0^{\infty } e^{-t} t^2 dt \Big(-\frac{2}{x^2t^2} \log \Big(\tan\frac{xt}{4}\Big)-\frac{1}{xt\sin\frac{xt}{2}}\Big).
\end{align}
Now $I_1(x) $ is simple to evaluate by performing the integral straightforwardedly and we get the answer as the following
\begin{align}\label{I1}
	I_1(x)&=\frac{1}{6} (-\log (x)-12 \zeta'(-1)-\log (2))-\frac{2 (-\log (x)+\gamma -1+2 \log (2))}{x^2}.
\end{align}
Using the integral representation of the Cosine Integral function, we have $I_2(x)$ to be
\begin{align}
I_2(x)=	\int_0^\infty e^{-t} t^2\sum_{n=0}^\infty (2n+1) \text{Ci}\Big((n+\frac{1}{2}) x t\Big)&=-\sum_{n=0}^\infty \int_0^\infty dt t^2 e^{-t} \int_{(n+\frac{1}{2}) xt}^\infty \frac{\cos z}{z} dz.
\end{align}
Interchanging the order of integrations in the above expression 
\begin{align}
	I_2(x)=-\sum_{n=0}^\infty (2n+1) \int_0^\infty dz \frac{\cos z}{z} \int_0^\frac{z}{(n+\frac{1}{2})x t} dt t^2 e^{-t}.
\end{align}
For the simplicity of the calculation we differentiate the above expression with respect to 
$x$ in the following step which is expressed in terms of polygamma functions and further it will be integrated to evaluate $I(x)$ up to a constant. The constant is also determined in the following.
\begin{align}
	\partial_x I_2(x)&={\rm Re}\sum_{n=0}^\infty\frac{2}{(n+\frac{1}{2})^2x^4}\int_0^\infty e^{-\Big(\frac{z}{(n+\frac{1}{2})x}+iz\Big)} z^2 dz\nonumber,\\
	&={\rm Re}\sum_{n=0}^\infty\frac{16 i (2 n+1)}{x (2 n x+x-2 i)^3}\nonumber,\\
	&={\rm Re}\Bigg[\frac{2 i \left(2 x \psi ^{(1)}\left(\frac{1}{2}-\frac{i}{x}\right)-i \psi ^{(2)}\left(\frac{1}{2}-\frac{i}{x}\right)\right)}{x^5}\Bigg],
\end{align}
where $\psi^{(n)}(x)$ denotes polygamma function of order $n$.
At large $x$ we have
\begin{align}
	\partial_x I_2(x)&=\frac{6 \psi ^{(2)}\left(\frac{1}{2}\right)}{x^5}+O(x^{-7}).
\end{align}
Now integrating the above expression we have 
\begin{align}\label{I2+cons}
	I_2(x)=-\frac{3 \psi ^{(2)}\left(\frac{1}{2}\right)}{2 x^4}+c''.
\end{align}
$c''$ is the constant of integration and it can be fixed by performing the integral $I_2(x)$ directly in the following manner
\begin{align}\label{I2 direct}
	I_2(x)&=\sum_{n=0}^\infty(2 n+1) \int_0^{\infty } e^{-t} t^2 \text{Ci}\left(\left(n+\frac{1}{2}\right) x t\right) \, dt\nonumber,\\
	&=\sum_{n=0}^\infty\frac{(2 n+1) \left(4 (2 n x+x)^2-\left((2 n x+x)^2+4\right)^2 \log \left(\frac{4}{(2 n x+x)^2}+1\right)+48\right)}{\left((2 n x+x)^2+4\right)^2}\nonumber,\\
	&=\sum _{n=0}^{\infty } \frac{24}{(2 n+1)^3 x^4}+O(x^{-6})\nonumber,\\
	&=-\frac{3 \psi ^{(2)}\left(\frac{1}{2}\right)}{2 x^4}+O(x^{-6}).
\end{align}
Thus comparing the equations \eqref{I2 direct} and \eqref{I2+cons} we find that 
\begin{align}
	c''=0.
\end{align}
Now we evaluate $I_3(x)$ in the following manner,
\begin{align}
	I_3(x)&=\int_{0}^{\infty} dt t^2 e^{-t}\Big(-\frac{2}{x^2t^2} \log \Big(\tan\frac{xt}{4}\Big)-\frac{1}{xt\sin\frac{xt}{2}}\Big),\\
	&=-\int_{0}^{\infty} dt  e^{-t} \frac{2}{x^2} \frac{d}{dt} \Big(t\log \tan \frac{xt}{4}\Big).
\end{align}
Integrating by parts
\begin{align}
	I_3(x)=-\frac{2}{x^2}\int_0^\infty dt t e^{-t}\log \tan \frac{xt}{4}.
\end{align}
We can rewrite the above expression in the following  manner
\begin{align}
	I_3(x)&=-\frac{2}{x^2} {\rm Re}\int_0^\infty dt t e^{-t} \log\Big[\frac{1-e^{-\frac{ixt}{2}}}{i(1+e^{-\frac{ixt}{2}})}\Big] \nonumber,\\
	&=\frac{2}{x^2}{\rm Re} \int_0^\infty dt t e^{-t} \sum_{n=1}^\infty \frac{e^{\frac{ixtn}{2}}+(-1)^n e^{\frac{ixtn}{2}}}{n}\nonumber,\\
	&=\frac{2}{x^2}{\rm Re} \sum _{n=0}^{\infty } \frac{2}{(2 n+1) \left(1+i \left(n+\frac{1}{2}\right) x\right)^2}\nonumber,\\
	&={\rm Re}\frac{2 \left(-x \psi ^{(0)}\left(\frac{1}{2}\right)+x \psi ^{(0)}\left(\frac{1}{2}-\frac{i}{x}\right)+i \psi ^{(1)}\left(\frac{1}{2}-\frac{i}{x}\right)\right)}{x^3}.
\end{align}
At large $x$ we have the following behaviour for $I_3(x)$
\begin{align}\label{I3}
	I_3(x)&=\frac{\psi ^{(2)}\left(\frac{1}{2}\right)}{x^4}+O(x^{-6}).
\end{align}
Finally combining \eqref{I1}, \eqref{I2 direct} and \eqref{I3} we show that the  apparently diverging sum    \eqref{original} converges to the following value at large $x$ after the reorganisation of the terms performed above
\begin{align}\label{I final}
	I=\frac{1}{6} (-\log (x)-12 \zeta'(-1)-\log (2))+O(\frac{\log x}{x^2}).
\end{align}
Note that contributions due to $I_2(x)$ and $I_3(x)$ appears in the higher order corrections in $\frac{1}{x}$ and are not important at large $x$. Thus we obtain the partition function for a massless  conformally coupled scalar on $S^1\times S^2$, putting \eqref{I final} into \eqref{lim m 0}, recalling $x=\frac{1}{mr}$,
\begin{align}
		\log Z\Big(0,\frac{\beta}{r}\Big)=&\frac{2 r^2 \zeta (3)}{\beta ^2}
	-\frac{1}{12}\log \frac{\beta}{r}-\zeta'(-1)-\frac{1}{12}\log 2
	+
	\frac{7 \beta ^2}{11520 r^2}+\frac{31 \beta ^4}{11612160 r^4}+\cdots\nonumber.\\
\end{align}
Observe that this agrees with the result obtained earlier  (C.22) of   \cite{Benjamin:2023qsc}
  for free conformal scalars on $S^1\times S^2$ including the constants which are independent of temperature.

	\bibliographystyle{JHEP}
	\bibliography{references} 	

\end{document}